\documentclass[%
 reprint,
superscriptaddress,
 amsmath,amssymb,
 aps,prl
]{revtex4-1}

\usepackage{dsfont}
\usepackage{graphicx}
\usepackage{dcolumn}
\usepackage{bm}
\usepackage[dvipsnames]{xcolor}
\usepackage{hyperref}
\hypersetup{
colorlinks=true,
	linkcolor=BlueViolet,
	citecolor=BlueViolet,
	urlcolor=BlueViolet,
}

\usepackage{ulem} 

\newcommand{\comment}[1]{}

\newcommand{\e}{\mathrm{e}}

\begin{document}
\definecolor{nrppurple}{RGB}{128,0,128}

\preprint{APS/123-QED}

\title{Circuit QED detection of induced two-fold anisotropic pairing in a hybrid superconductor-ferromagnet bilayer}

\author{C.~G.~L.~B\o{}ttcher}
\thanks{These authors contributed equally to this work. \\ Email: charlotte.boettcher@yale.edu\\
C.G.L.B. present address: Department of Applied Physics, Yale University, New Haven, CT, 06520, USA.\\}
\affiliation{Department of Physics, Harvard University, Cambridge, MA 02138, USA}
\author{N.~R.~Poniatowski}
\thanks{These authors contributed equally to this work. \\ Email: charlotte.boettcher@yale.edu\\
C.G.L.B. present address: Department of Applied Physics, Yale University, New Haven, CT, 06520, USA.\\}
\affiliation{Department of Physics, Harvard University, Cambridge, MA 02138, USA}
\author{A.~Grankin}
\affiliation{Joint Quantum Institute, Department of Physics, University
of Maryland, College Park, MD 20742, USA}
\author{M.~E.~Wesson}
\affiliation{Harvard John A. Paulson School of Engineering and Applied Sciences, Harvard University, MA 02138, Cambridge, USA}
\author{Z.~Yan}
\affiliation{Department of Physics, Harvard University, Cambridge, MA 02138, USA}
\author{U.~Vool}
\affiliation{Department of Physics, Harvard University, Cambridge, MA 02138, USA}
\affiliation{Max Planck Institute for Chemical Physics of Solids, Dresden 01187, Germany}
\author{V.~M.~Galitski}
\affiliation{Joint Quantum Institute, Department of Physics, University
of Maryland, College Park, MD 20742, USA}
\affiliation{Center for Computational Quantum Physics, The Flatiron Institute, New York, NY 10010, United States}
\author{A.~Yacoby}
\affiliation{Department of Physics, Harvard University, Cambridge, MA 02138, USA}
\affiliation{Harvard John A. Paulson School of Engineering and Applied Sciences, Harvard University, MA 02138, Cambridge, USA}

\date{\today}

\begin{abstract}
    Hybrid systems represent one of the frontiers in the study of unconventional superconductivity and are a promising platform to realize topological superconducting states. Owing to their mesoscopic dimensions, these materials are challenging to probe using many conventional measurement techniques, and require new experimental probes to successfully characterize. In this work, we develop a  probe that enables us to measure the superfluid density of micron-size superconductors using microwave techniques drawn from circuit quantum electrodynamics (cQED). We apply this technique to a paradigmatic hybrid system, the superconductor/ferromagnet bilayer, and find that the proximity-induced superfluid density is two-fold anisotropic within the plane of the sample and exhibits power law temperature-scaling which is indicative of a nodal superconducting state. These experimental results are consistent with the theoretically predicted signatures of induced triplet pairing with a nodal $p$-wave order parameter. Moreover, we unexpectedly observe drastic modifications to the microwave response at frequencies near the ferromagnetic resonance, suggesting a coupling between the spin dynamics and induced superconducting order in the ferromagnetic layer.  Our results offer new insights into the unconventional superconducting states induced in superconductor/ferromagnet heterostructures and simultaneously establish a new avenue for the study of fragile unconventional superconductivity in low-dimensional materials such as van der Waals heterostructures.  
\end{abstract}

\maketitle

Heterostructures constructed from superconductors and other materials (e.g. semiconductors, ferromagnets, and topological materials) offer a rich platform to realize unconventional superconducting states via proximity effects. In these hybrid systems, the coupling between distinct materials leads to the formation of emergent phases that feature new physical properties that are otherwise absent in the isolated constituents. These include topological superconducting phases \cite{Sato_review} hosting non-Abelian excitations, as well as states supporting spin-triplet pairing \cite{sc-spintronics}, both of which have potential applications for quantum computing technology \cite{sds-rmp}. Given the extreme scarcity of naturally occurring topological \cite{volovik-3heb,jim-upt3,ins-upt3,upt3-rmp} or spin-triplet superconductors \cite{u-sc,ute2,cubise,bbg-sc}, hybrid systems are an invaluable resource to realize these exotic superconducting states \cite{lsd-majorana,Alicea_majorana,bergeret-1,Ren2019}.

The archetypal superconducting hybrid system is the superconductor-ferromagnet (S/F) bilayer, where spin-triplet superconductivity can be induced in the ferromagnet due to the combined effect of the exchange field and superconducting proximity effect \cite{buzdin-rmp,odd-triplet-rmp}. These systems are well-studied for their numerous applications for superconducting spintronics \cite{sc-spintronics,Eschrig-review}, as well as their potential use as a platform for topological quantum computing. However, while S/F heterostructures have been extensively studied using transport techniques, direct probes of the induced superconducting order have been lacking. This disparity lies in the fact that the mesoscopic nature of the induced superconducting state, which is tightly confined to the S/F interface and exists over nanometer-scale distances, renders most well-developed techniques in the study of bulk unconventional superconductors (e.g. heat capacity, penetration depth, or neutron scattering measurements) challenging to apply. Although there has been promising recent work applying conventional techniques to superconductor heterostructures \cite{sf-twocoil,anlage-proximity,para-meissner-musr}, new experimental probes are required to enable the direct study of induced unconventional superconducting states in hybrid superconducting systems \cite{higginbotham-prl,spincurrent}. 

In this work, we employ an on-chip superconducting microwave resonator as a sensitive probe of a micron-scale S/F bilayer. Resonator circuits allow for the creation and control of highly localized electromagnetic fields, enabling one to attain strong coupling even to micron-scale samples \cite{yig-3d,yig-3d-2,yig-2d,nb-py-prl1,nb-py-prl2}. By integrating a S/F bilayer into our resonator circuit, one can probe the inductive response of the bilayer which, as will be discussed at length below, is a direct manifestation of the induced superfluid density in the hybrid system. The temperature dependence of the superfluid density directly reflects the underlying pairing symmetry of the superconducting state, and has been employed with great success to gain insight into bulk superconductors \cite{Prozorov}. However, temperature-dependent superfluid density measurements have yet to be applied to the study of superconductivity in mesoscopic systems. 

When a metallic ferromagnet is placed into contact with a conventional $s$-wave superconductor, the strong exchange field in the ferromagnet depairs the spin-singlet Cooper pairs inherited from the superconductor and suppresses induced singlet superconductivity \cite{eugene-sf}. However, interfacial spin-orbit coupling (which is generically present) or magnetic inhomogeneities can flip the spin of an electron as it tunnels across the S/F interface and convert singlet pairs into spin-triplet Cooper pairs which can survive in the ferromagnet, leading to the formation of a mini-gap in the majority-spin band in the ferromagnet \cite{sf-soc,takei_SF}. To satisfy fermionic antisymmetry, these triplet pairs must either have an odd-parity (e.g. $p$-wave) orbital structure or an ``odd-frequency’’ pairing structure, with the pairing correlations being antisymmetric with respect to time \cite{odd-triplet-rmp,odd-w-rmp}. Although the presence of triplet pairs has been indirectly inferred from the persistence of long-range supercurrents in long S/F/S Josephson junctions \cite{sfs-jj-1,sfs-jj-2,sfs-jj-3}, the detailed symmetry of the induced pairing is not yet well understood.

 \begin{figure*}
	\includegraphics[width= 6.5 in]{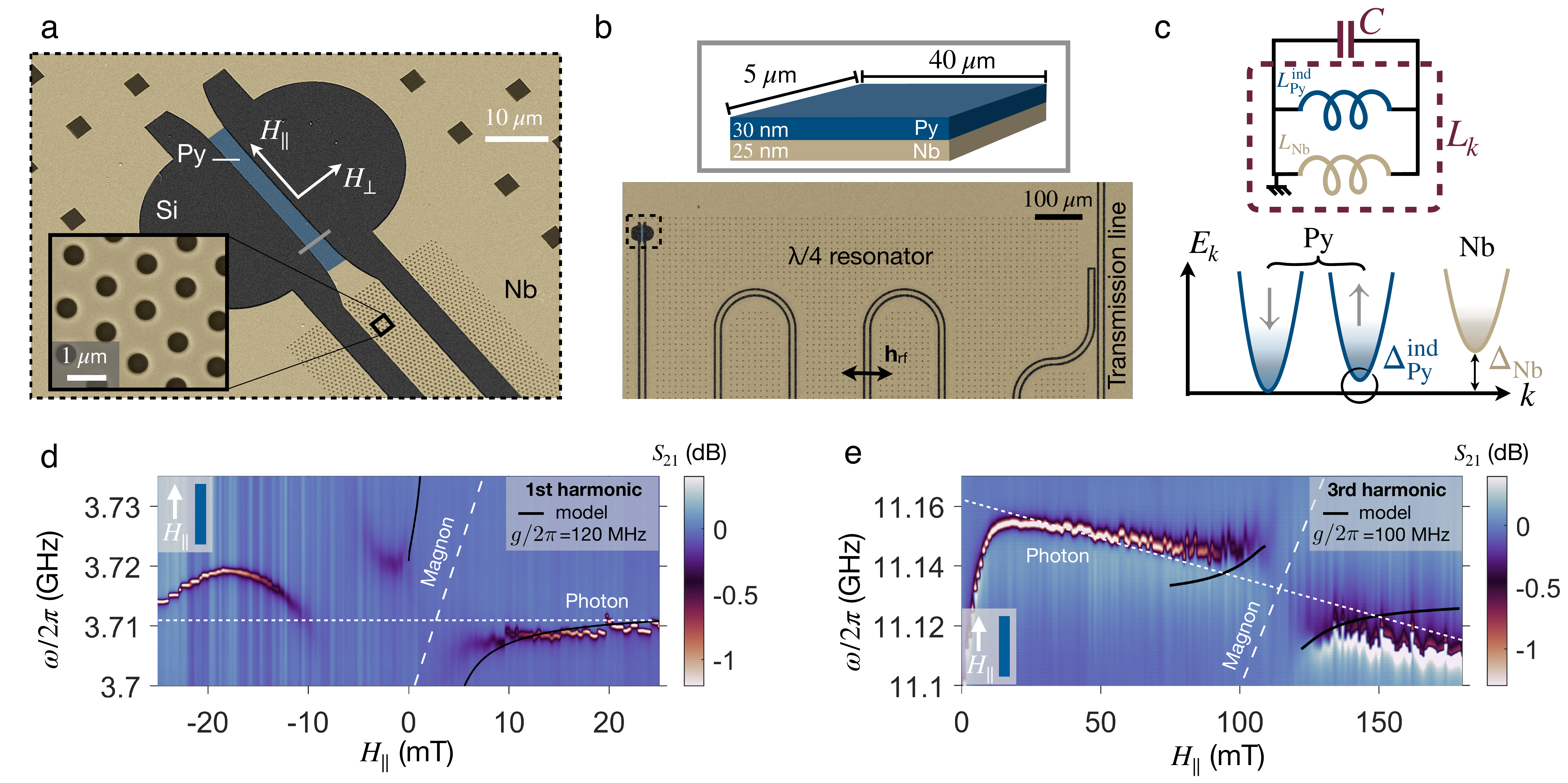} 
	\caption{\textbf{Device geometry and ferromagnetic resonance}. \textbf{a.} False-colored scanning electron micrograph of the superconductor/ferromagnet (S/F) bilayer composed of a 30 nm permalloy film deposited directly on top of a 25 nm thick Nb film, as shown in the cross-section in panel \textbf{b}. The bilayer is integrated into a quarter-wavelength coplanar resonator patterned into the Nb film, shown in an optical micrograph in panel \textbf{b}. The resonator is capacitively coupled to a transmission line and is perforated with artificial flux-pinning holes (inset of panel \textbf{a}.) to improve the resonator performance in magnetic fields. \textbf{c.} Top: At microwave frequencies, the bilayer response can be treated as a circuit of two parallel inductors, corresponding to the kinetic inductances associated with the bulk Nb superfluid density ($L_{\text{Nb}} \sim 1/n_s^{\text{Nb}}$) and the induced superfluid density in the bilayer. Bottom: As a result of their direct contact, the Nb is able to proximity induce superconductivity in the Py strip, leading to the formation of a mini-gap $\Delta_{\text{Py}}$ in the majority spin band. \textbf{d.} Transmission $S_{21}$ across the circuit as a function of in-plane magnetic field $\mu_0 H_\parallel$ oriented along the length of the Py stripe. When the resonator frequency is tuned to the ferromagnetic resonance frequency of the Kittel magnons in the Py, an anti-crossing is observed in the resonator spectrum, where the cavity photons hybridize with the FMR mode to form cavity magnon-polaritons. Note that given the low field $\mu_0 H_\parallel \approx 7$ mT associated with the FMR at this frequency, the magnon-photon hybridization leads to substantial damping of the photon mode even at zero field. The black lines is an overlay of the modelled spectral function of coupled harmonic oscillators (see Supplemental Information), which allow us to extract an effective coupling strength $g/2\pi = 120$ MHz between the resonator photons and Py magnons. \textbf{e.} Transmission spectrum at the third harmonic of the resonator. Anti-crossings associated with magnon-polariton modes are again observed, now at a higher field $\mu_0 H_\parallel \approx 120$ mT, at which the FMR crosses the third harmonic frequency $\approx 11$ GHz. Fitting the transmission spectrum (black lines) yields a similar coupling $g/2\pi = 100$ MHz to that observed at the first harmonic. The broad shoulder on the left-hand side of both sweeps is due to hysteretic effects related to trapped flux in the superconducting resonator.} 
 \label{schematic-fig}
\end{figure*}

To directly address the induced superconducting state in an S/F bilayer requires a probe that is amenable to the small spatial size of typical devices (with nanometer to micron scale dimensions), as well as the ability to selectively address the weak induced superconducting state that exists in parallel with the intrinsic bulk superconductivity of the superconducting layer. To achieve both of these requirements, we employ an on-chip superconducting coplanar waveguide resonator, which has been extensively developed as a part of the circuit quantum electrodynamics (cQED) architecture for superconducting quantum information devices \cite{blais-rmp}. The resonator is fabricated from a 25 nm-thick Nb film perforated with flux pinning holes to maximize performance in external magnetic fields \cite{leo-fluxholes}, in a quarter-wavelength configuration with one end of the resonator shorted to ground and the other open (Fig. \ref{schematic-fig}a,b). These Nb resonators are designed to have resonance frequencies $\omega_{\text{r}}/2\pi = 4-7$ GHz and attain quality factors of $Q \approx 350,000$ at our base operating temperature of $T \approx 55$ mK, enabling high sensitivity in our measurements.

To study the superconducting state of an S/F bilayer, we deposit a 30 nm thick permalloy (Py) stripe directly on top of the Nb center conductor where the resonator is shorted to ground, forming a Nb/Py bilayer that is situated at a current antinode of the circuit as shown in Fig. \ref{schematic-fig}a. Because the current is concentrated at the location of the bilayer, the resonator response is dominated by the properties of the S/F subsystem. One manifestation of the strong coupling between the Nb resonator and Py stripe is the drastic reduction in the quality factor of the Py-loaded resonator to $Q \approx 7,000$. 

Moreover, spectroscopy of the hanger style resonator, interrogated via the transmission $S_{21}$ across a capacitively coupled transmission line (Fig. \ref{schematic-fig}b), allows one to directly probe the microwave dynamics of the ferromagnet. In particular, by applying an in-plane magnetic field $H_\parallel$ along the length of the Py stripe, we may tune the ferromagnetic resonance (FMR) frequency which follows the Kittel law $\omega_{\text{m}}(H_\parallel) = \gamma \sqrt{(H_\parallel + M_s) H_\parallel}$, where $\gamma$ is the gyromagnetic constant and $\mu_0 M_s \approx 1.2$ T is the saturation magnetization of Py \cite{kittel,sw-book}. When the FMR frequency is brought to coincide with the frequency of the resonator mode (which is only weakly affected by small in-plane fields due to pair-breaking effects), we observe clear anti-crossings in the resonator spectrum associated with the formation of magnon-polaritons. These anti-crossings are observed at both low fields $\mu_0 H_\parallel \approx 7$ mT when the FMR intersects the fundamental frequency of the resonator, as well at higher fields $\mu_0 H_\parallel \approx 110$ mT when the FMR crosses the third-harmonic of the resonator (a quarter-wavelength resonator exhibits only odd harmonics), as shown in Fig. \ref{schematic-fig}d,e. In both cases, we can fit the resonator spectrum to the spectral function of two coupled harmonic oscillators, and extract an effective coupling strength $g/2\pi \approx 100$ MHz between the resonator and ferromagnetic resonance mode (see Supplemental Information for details). From the dimensions of the Py stripe, this corresponds to a coupling strength of 150 Hz/spin, which drastically exceeds the coupling strengths reported in prior works \cite{nb-py-prl1,nb-py-prl2} on Py/Nb hybrid circuits, where the Py stripe was separated from the superconductor with an insulating layer to prevent any degradation in $Q$ from the inverse proximity effect. In contrast, our devices feature a direct interface between the superconductor and ferromagnet, enabling proximity effects and the possibility of new dynamics generated by the interplay between the order parameters in each layer of the hybrid S/F system.

At microwave frequencies, a superconductor behaves as an inductive element characterized by a ``kinetic inductance'' arising from the superfluid response \cite{degennes-book}. This inductance is fundamentally related to the density $n_s$ of superfluid carriers as $L_K = (m/n_s e^2)(\ell/s)$ for a superconducting wire of length $\ell$ and cross-section $s$. Notably, the kinetic inductance is large for fragile or dilute superconductors with a small superfluid density, as well as for thin systems with small cross-sections. Both of these features make kinetic inductance measurements especially favorable for probing weak and low-dimensional superconductors, such as the proximity-induced superconducting state in an S/F bilayer. The kinetic inductance is reflected in the resonant frequency $2\pi f = 1/\sqrt{(L_\text{g} + L_K)C}$, where $L_\text{g}$ and $C$ are the geometric inductance and capacitance of the circuit, respectively. When the system is weakly perturbed by changing an external parameter such as the temperature or applied magnetic field (under the reasonable assumption that $L_{\text{g}}$ and $C$ are constant), the frequency shift of the resonator is directly proportional to the change in the kinetic inductance or, equivalently, in the superfluid density
\begin{equation}
    \frac{\delta f}{f_0} \approx -\frac{1}{2} \frac{\delta L_K}{L_{K,0}} \approx \frac{\kappa}{2} \frac{\delta n_s}{n_{s,0}} \, , 
\end{equation}
where we have assumed that the frequency shift is small compared to the resonance frequency $f_0$ at our base operating temperature of 55 mK such that $\delta f/f_0 \ll 1$, and have introduced the kinetic inductance fraction $\kappa = L_{K,0}/(L_{\text{g}} + L_{K,0})$. Thus, by studying the evolution of the resonator frequency with temperature or magnetic field, we are able to sensitively measure the changes in the superfluid density of the S/F bilayer, offering a direct probe of the induced superconducting order.

In fact, superfluid density measurements have proven to be an essential tool in the study of unconventional superconductivity \cite{Prozorov,sigrist-rmp}. In conventional fully gapped superconductors, the superfluid density exhibits a thermally activated temperature dependence $\delta n_s(T)/n_{s,0} \equiv \left[n_s(T) - n_s(0)\right]/n_s(0) \sim \mathrm{e}^{-\Delta/T}/\sqrt{T}$ where $\Delta$ is the superconducting energy gap. In contrast, unconventional superconductors with nodal order parameters host  low-lying quasiparticles residing at the gap nodes, leading to a power law dependence of the superfluid density $\delta n_s(T)/n_{s,0} \sim T^n$, where the exponent $n$ depends on spatial dimensionality, the dimensionality of the nodes, and the degree of disorder in the system \cite{Prozorov,sigrist-rmp}. As a consequence of this distinction between activated and power law temperature dependence, superfluid density measurements have become a standard, powerful tool in assessing the gap topology of bulk superconductors. 

 \begin{figure}[t]
	\includegraphics[width= 3.5 in]{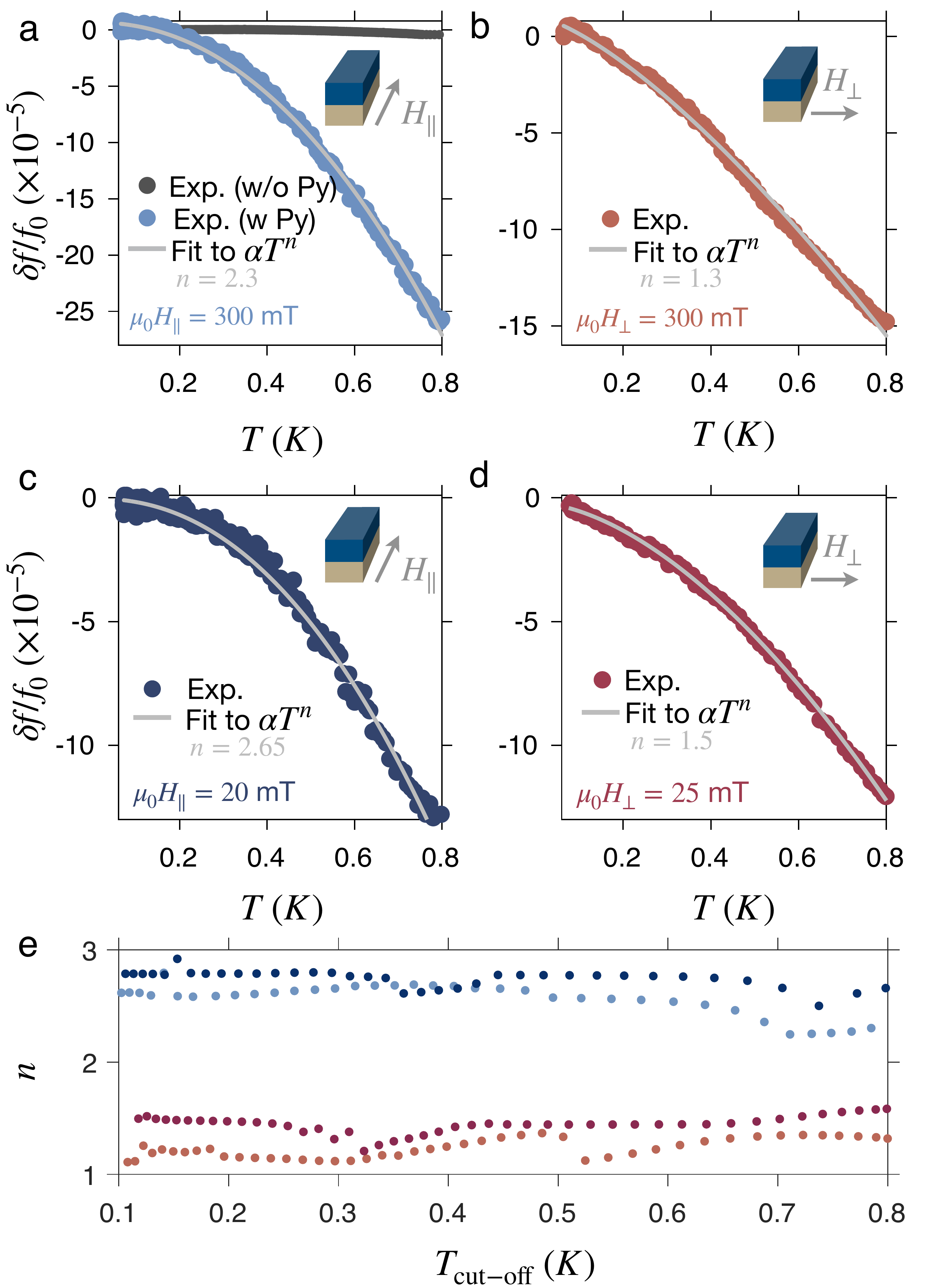} 
	\caption{\textbf{Anisotropic temperature dependence of inductance}. \textbf{a.} Shift in resonance frequency $\delta f/f_0 = [f(T,H) - f(55 \; \text{mK},H)]/f(55 \; \text{mK},H)$ in an in-plane field $\mu_0 H_\parallel = 300$ mT oriented along the length of the Py stripe, as illustrated in the inset. The comparatively negligible temperature dependence of the resonance frequency of a bare Nb resonator (without a Py stripe) is shown for comparison. \textbf{b.} Shift in the resonance frequency in an in-plane field $\mu_0 H_\perp = 300$ mT oriented perpendicular to the length of the Py stripe. \textbf{c.} Frequency shift in an in-plane field $\mu_0 H_\parallel = 20$ mT; \textbf{d.} Frequency shift in an in-plane field $\mu_0 H_\perp = 25$ mT. In all plots, the grey line is a fit of the data over the full temperature range to the power-law dependence $\delta f/f_0 = \alpha T^n$, with $\alpha$ and $n$ fitting parameters. \textbf{e.} Extracted temperature-scaling exponent $n$ as a function of the upper cutoff of the temperature range over which the data is fit, for the data in each panel \textbf{b}-\textbf{d}. Irrespective of the details of the fit procedure, the scaling exponents for fields parallel and perpendicular to the stripe are distinct.}
 \label{anisotropy-fig}
\end{figure}

We may simplistically imagine that the microwave response of the bilayer can be described as that of two parallel inductors, as illustrated in Fig. \ref{schematic-fig}c: one corresponding to the kinetic inductance of the induced superconducting state in the Py, and the other corresponding to the bulk superfluid density of the Nb film below. We will focus on the low-temperature regime $T \lesssim 800$ mK in our measurements, well below the critical temperature $T_c^{\text{Nb}} \approx 8$ K, such that the kinetic inductance of the Nb film is effectively frozen out and equal to its zero-temperature value. Experimentally, as shown in Fig. \ref{anisotropy-fig}a, the resonance frequency of bare Nb resonators exhibits very little temperature dependence in this range, with $\delta f/f_0 \sim 10^{-6}$, consistent with this assumption. Moreover, we have further validated this technique by measuring the superfluid density of a small micron-scale Al film inserted at the end of the resonator (see Supplemental Information), which leads to an activated temperature-dependence of the resonance frequency with a rate consistent with the gap of Al. Thus, we can attribute the temperature-dependent changes studied below to the microwave response of the S/F bilayer. 


We may begin by studying the response of the hybrid resonator in an applied in-plane magnetic field so that the ferromagnetic resonance is detuned to be far above our operating frequency, which in this case is $\omega_r/2\pi \approx 7$ GHz. The system's behavior when the ferromagnetic resonance is near the resonator frequency, and the cavity mode takes on the character of a magnon-polariton, will be discussed later. 
In Fig. \ref{anisotropy-fig}a, we present the temperature dependence of the fundamental resonant frequency of the hybrid superconductor-ferromagnet circuit in an in-plane magnetic field of $\mu_0 H_\parallel = 300$ mT oriented along the length of the Py stripe. Notably, the temperature dependence is manifestly non-exponential, in contrast to the expectation for a conventional fully gapped superconductor. Fitting the temperature-dependent frequency shift to a simple power law, $\delta f/f_0 = \alpha \, T^n$, we find an exponent of $n = 2.3$. The overall magnitude $\alpha$ of the frequency shift is determined by several non-universal factors, and we will primarily focus on the exponent $n$ throughout this work (see the Supplemental Information for further discussion). In this measurement configuration, the external magnetic field $\mu_0 H_\parallel$ is applied parallel to the microwave current flowing in the resonator, as shown in the inset. 

In Fig. \ref{anisotropy-fig}b, we instead apply the field perpendicular to the current, and present the temperature dependence of $\delta f/f_0$ at $\mu_0 H_\perp = 300$ mT. We again find a power-law, rather than exponential, temperature dependence with a different, faster exponent $n = 1.3$ compared to the $H_\parallel$ configuration. That is, we observe a two-fold anisotropy in the temperature scaling of the hybrid resonator frequency, and by extension of the kinetic inductance of the S/F bilayer. 

We may also perform measurements at lower magnetic fields, and probe the temperature dependence of the resonance below the FMR frequency. In Fig. \ref{anisotropy-fig}c,d we present $\delta f/f_0$ traces for $\mu_0 H_\parallel = 20$ mT and $\mu_0 H_\perp = 25$ mT, where we again see power-law temperature dependences in both cases. Further, we again find a two-fold anisotropy in the exponent $n$, with $n = 2.65$ in the parallel configuration and $n = 1.5$ in the perpendicular configuration. Thus, we find that the temperature-dependent response of the S/F bilayer is qualitatively unchanged by the magnitude of the applied magnetic field. 

The temperature in our dilution refrigerator is only stable below $\sim 800$ mK, constraining the accessible temperature range for our measurements. To ensure that our results are independent of this upper limit, we may restrict the fits of the temperature-dependent resonance frequencies to progressively lower temperatures, and extract the scaling exponents $n$ for each field orientation for different values of the upper cutoff of the fitting range, $T_{\text{cut-off}}$. We plot the extracted exponents $n$ as a function of $T_{\text{cut-off}}$ in Fig. \ref{anisotropy-fig}e, where we see that the scaling exponents for parallel vs. perpendicular field orientations are clearly distinct independent of the fitting range, emphasizing the robustness of the observed scaling anisotropy.

It is natural to attribute the temperature dependence of the induced superfluid density, manifested in the shift of the hybrid S/F resonance, to the thermal excitations above the proximity-induced mini-gap. In particular, we note that the features we observe occur on temperature scales on the order of tens to hundreds of milliKelvin, which is substantially smaller than the energy scales associated with either the superconductor (with a critical temperature $\approx 8$ K) or ferromagnet (with a Curie temperature $\approx 500$ K) independently. This strongly suggests that the physics underlying the observed temperature-dependent response arises due to the low-energy coupling between the two states, e.g. from a proximity-induced superconducting state. 

In general, a variety of superconducting correlations with different spin and orbital symmetries are generated at the S/F interface \cite{buzdin-rmp,odd-triplet-rmp,Eschrig-review}. Typically, however, only the correlations which can persist over long distances (such as the odd-frequency triplet state) into the ferromagnet contribute meaningfully in traditional transport experiments, and hence have been the principal focus of theoretical study. Nonetheless, other superconducting correlations are always present, albeit potentially confined to the interface over atomic-scale distances and thus challenging to detect using conventional probes. 

Our observation of a power-law, rather than activated, temperature-dependence of the superfluid density suggests that we are coupling to a nodal, rather than fully-gapped, induced superconducting state. Such an anisotropic state would not be protected by Anderson's theorem and thus susceptible to pair-breaking from impurity scattering, and consequently would be confined to within a mean free path of the S/F interface. The possibility of our experiment to detect such a weak state lies in the fact that we measure \textit{changes} in the kinetic inductance, and thus are primarily sensitive to the lowest-lying thermally excited quasiparticles and the most fragile superconducting states, as opposed to being immediately shunted by the fully-gapped superconducting state. Moreover, the lateral geometry of the bilayer integrated into our superconducting circuit enables even states localized to the S/F interface to contribute to the inductive response. 

In general, the superfluid density is a tensor quantity that can have two distinct components in a (quasi-) two-dimensional system \cite{Prozorov,leggett-rmp,3he-book}. Moreover, the superfluid density in a nodal superconductor can display different temperature scalings depending on the relative orientation between the current and nodal direction \cite{3he-book}. Intuitively, one can imagine that the gapless quasiparticle states residing near the gap nodes are most efficiently excited when the current is aligned along the nodal direction, leading to a temperature scaling $\delta n_s \sim T$ that reflects the linear dispersion of the nodal quasiparticles. In contrast, when the current is aligned along the anti-nodal direction, nodal quasiparticles are less efficiently excited, leading to a slower temperature dependence $\delta n_s \sim T^n$ with $n > 1$. In this case, the precise power law dependence of the superfluid density is determined by the microscopic details of the system (e.g. spatial dimensionality, co-dimension of the gap nodes, disorder, etc.).

Thus, our finding of a two-fold anisotropic power-law scaling of the superfluid density strongly constrains the possible superconducting states detected in our experiment. In particular, the two-fold anisotropy is only consistent with an induced order parameter with a $p$-wave orbital symmetry. Moreover, the power-law dependence of the superfluid density implies that the induced state is nodal, and that by applying the dc magnetic field parallel or perpendicular to the microwave current, we are able to selectively address a nodal and anti-nodal orientation of the $p$-wave order parameter.

\begin{figure}
\begin{centering}
\includegraphics[scale=0.3]{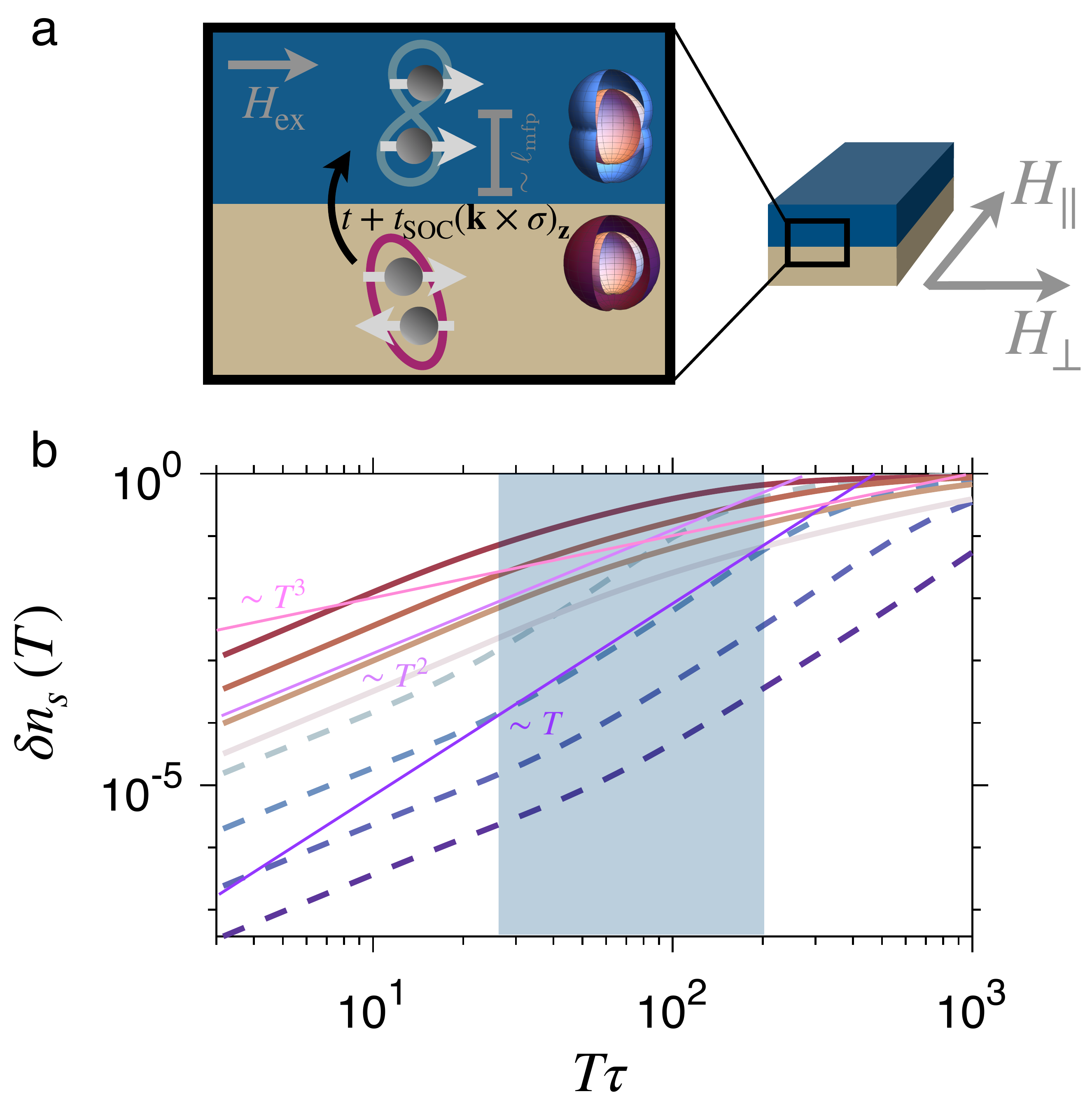}
\par\end{centering}
\caption{\textbf{Superfluid density for a disorder nodal $p$-wave state. } \textbf{a.} Illustration of the S/F bilayer with the in-plane field directions $H_\parallel$, $H_\perp$ indicated. The cross-section schematically depicts how interfacial spin-orbit coupling can convert isotropic spin-singlet pairs in the Nb layer into spin-triplet $p$-wave pairs in the ferromagnet, which can survive into the ferromagnet over lengths scales on the order of the electronic mean free path. \textbf{b.} Superfluid density $\delta n_{s}\left(T\right)=n_{s}\left(T\right)-n_{s}\left(T=0\right)$
as a function of temperature for $\Delta\tau\approx5\times10^{2}$,
$\Delta\tau\approx10^{3}$, $\Delta\tau\approx3\times10^{3}$, $\Delta\tau\approx6\times10^{3}$, where $\Delta$ is the triplet gap. Solid and dashed
curves correspond to the response along and transverse to the nodes
of the superconducting gap, respectively, where darker colors correspond to higher $\tau$. Lines corresponding to temperature scalings of $T,T^2,$ and $T^3$ are included in purple/pink as guides to the eye. The blue-shaded region indicates the range of parameter space compatible with the experimental results.}

\label{Fig_theor}

\end{figure}

To inform our experimental findings, we now construct a phenomenological model for the induced superfluid density in the S/F bilayer.
We consider a bilayer system consisting of an $s$-wave superconductor
and a ferromagnet with an in-plane magnetization oriented along ${\bf H}_{ex}$.
The inter-layer tunneling is assumed to have a spin-independent component, $t$,
as well a component with the Rashba spin-orbit texture $\propto t_{\text{soc}}\left({\bf k}\times\sigma\right)_{z}$
where $\sigma$ is the electron spin and ${\bf k}$ is the in-plane
electron momentum. The spin-orbit coupling at the interface arises due to
the inversion symmetry breaking, and, as was shown in \cite{takei_SF},
can generate chiral $p_{x}+ip_{y}$ superconducting correlations in
the ferromagnet when the magnetization is oriented out of the plane of the sample. Here, we take the Zeeman field to lie in an in-plane orientation,
which gives rise to  a nodal $p$-wave  order parameter. 
In the Supplemental Information, we use this model to derive
the effective $p$-wave order parameter for the majority spin component
of the ferromagnet, which is shown to have the form $\Delta_{{\bf k}}=\Delta_{t}\cos\theta$,
where $\theta$ is the angle between ${\bf k}$ and ${\bf H}_{ex}$,
and $\Delta_{t}$ is the amplitude of the triplet order parameter.
Within the mean-field approximation, the Meissner kernel at temperature $T$ can be found to be \cite{hirschfeld-disorder}
\begin{equation}
\delta K_{i,j}=\frac{-2e^{2}}{c}\int\limits_{0}^{\infty}d\epsilon n_{F}\left(\epsilon\right)\left\langle \Re\frac{ {\bf  v}_{i}{\bf v}_{j} \Delta_{{\bf k}}^{2}}{\left[\left(\epsilon-\Sigma\right)^{2}-\Delta_{{\bf k}}^{2}\right]^{3/2}}\right\rangle _{FS},\label{eq:Kij}
\end{equation}
where $\langle\ldots\rangle_{FS}$ denotes a Fermi surface average,
${\bf v}_{i}$ is the Fermi velocity, $n_{F}\left(\epsilon\right)$
is the Fermi distribution, $\delta K_{i,j}=K_{i,j}\left(T\right)-K_{i,j}\left(0\right)$,
$\Sigma\left(\epsilon\right)$ is the diagonal component of the self-energy
which we evaluate within the strong-scattering self-consistent $T-$matrix
approximation $\hat{\Sigma}\left(\epsilon\right)=\tau^{-1}/\sum_{k}\hat{G}_{k}$,
where $\tau$ is the scattering time
and $\hat{G}$ is the Nambu electron Green's function. We note
that vertex corrections must be included due to the anisotropy of the order parameter,
as discussed in the Supplemental Information. At finite disorder,
the low-temperature scaling of the Meissner kernel is quadratic $\delta K_{i,i}\sim T^{2}$.
At higher temperatures the response scales as $T$ and $T^{3}$ when
probed along the nodal and antinodal directions of the superconducting order parameter, respectively.
For a general temperature and disorder scattering time $\tau$, the
Meissner response can be evaluated numerically. The result for the
anisotropic superfluid density defined as $n_{s}=K_{i,j}c/e^{2}$
is shown in Fig.~\ref{Fig_theor}, where we find that the temperature scaling continuously evolves from a quasi-isotropic $\sim T^2$ dependence at strong disorder to a strongly anisotropic $T$/$T^3$ dependence along the nodal/antinodal direction in the clean or high-temperature limit. Notably, the experimentally observed temperature scaling along the two directions is compatible with this theory over a wide region of parameter space, highlighted in blue in Fig. \ref{Fig_theor}. 

To intuitively understand the origin of these power laws, we recall that in a clean superconductor with line nodes, one expects that the component of the superfluid density along the nodal direction scales linearly with temperature, reflecting the linear dispersion of the low-lying quasiparticles as discussed above. However, the introduction of weak non-magnetic disorder gives rise to low-lying impurity states which ``fill in'' the node, leading to a finite quasiparticle density of states at low energies, manifested as a quadratic temperature dependence of the superfluid density at low temperatures \cite{hirschfeld-disorder}. Above the energy scale $T^\star$ of these impurity states (which is set by the superconducting gap and impurity scattering rate), the usual linear-in-temperature scaling is recovered. In fact, such a quadratic-to-linear crossover has been extensively used to successfully describe superfluid density measurements of cuprate superconductors with varying degrees of disorder. In the language of temperature-scaling exponents, this quadratic-to-linear crossover translates to intermediate scaling exponents $1 < n < 2$ in the nodal direction (as observed experimentally), where the precise value of $n$ varies continuously with the degree of disorder. Similarly, one expects $2 < n < 3$ in the antinodal direction, which is again consistent with the experimental results. 

\begin{figure}[t]
	\includegraphics[width= 3.5 in]{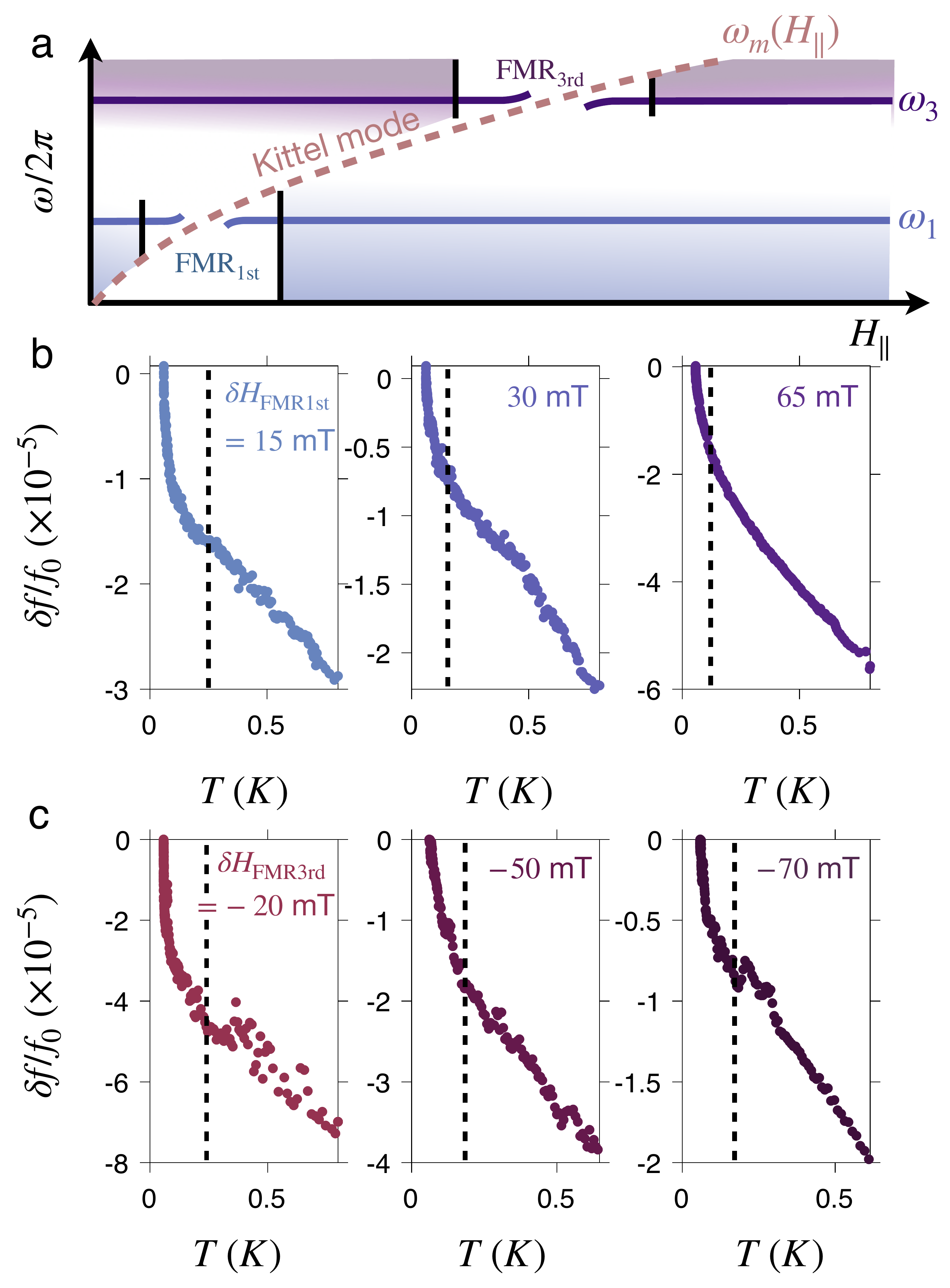} 
	\caption{\textbf{Temperature scaling near the ferromagnetic resonance}. \textbf{a.} Schematic illustration of the first and third harmonic modes of the resonator and the evolution of the ferromagnetic resonance (Kittel) mode frequency with in-plane magnetic field. \textbf{b.} Temperature dependence of the first harmonic resonator frequency at fields above the ferromagnetic resonance field, i.e. $\delta H_{\text{FMR,1st}} = H - H_{\text{FMR,1st}} >0$. Progressively steeper upturns in the temperature dependence are observed as the ferromagnetic resonance field is approached. \textbf{c.} Temperature dependence of the third harmonic resonator frequency at fields below the ferromagnetic resonance field, $\delta H_{\text{FMR,3rd}} < 0$. The steepness of the upturns again scales with the proximity to the ferromagnetic resonance field. Dashed lines are guides the eye that mark the approximate temperature at which the upturns onset.}
 \label{fmr-fig}
\end{figure}

So far, we have focused on the temperature-dependent response of the hybrid S/F resonator subjected to in-plane magnetic fields such that the resonator is far detuned from the ferromagnetic resonance frequency. If we perform the same measurements at fields where the resonator frequency is near the FMR frequency, we observe strikingly different behavior as illustrated in Fig. \ref{fmr-fig}. Namely, we observe a sharp ``upturn’’ in the resonance frequency as the temperature is lowered, which can be described as a nearly divergent power law scaling $\delta f/f_0 \sim T^n$ with $n < 1$ at low temperatures. By comparing the response of the first and third harmonics, which intersect the FMR at different magnetic fields, we can confirm that these upturn features track with the proximity to the FMR field (i.e. the field $H_{\text{FMR}}$ such that  $\omega_m(H_{\text{FMR}}) = \omega_r$) as opposed to the magnitude of the in-plane magnetic field itself. These upturns become increasingly sharp as the FMR field is approached, and weaker upturns persist over a relatively wide field range, on the order of 100 mT, away from the FMR field. 
The exact field range over which the upturns persist is device-dependent, but in all cases the upturns track with the FMR frequency. 

The appearance of these low-temperature upturns, which manifest on temperature scales far lower than the relevant scales in either the superconductor or ferromagnet independently, are again indicative of strong interactions and hybridization between the two subsystems. However, on account of the unusual, seemingly divergent, temperature-scaling exponent $n$ in this regime, it is unclear whether the temperature dependence of the resonance frequency near the FMR can be simply attributed to changes in the superfluid density of the S/F bilayer. We also note that qualitatively similar upturns, and history-dependent artifacts presumably related to trapped magnetic flux, are occasionally observed in the temperature-dependence of bare Nb resonators after repeated magnetic field cycling (as elaborated on in the Supplemental Information). In contrast, the upturns observed in the S/F devices near the FMR are a reproducible feature of the phenomenology of these devices. 

Regardless of whether the low-temperature upturns can be associated with the induced superfluid density, these unusual features clearly reflect a non-trivial low-energy coupling between the superconductor and ferromagnet subsystems. The origin of these upturns is yet to be theoretically understood and necessitates further study of the coupled dynamics of S/F heterostructures. 

Altogether, our kinetic inductance measurement technique has enabled us to access previously inaccessible aspects of the physics of S/F heterostructures. We are able to sensitively couple to fragile sub-dominant induced superconducting orders, beyond the usual long-range triplet states that typically dominate the transport response of S/F systems. Our work thus establishes kinetic inductance techniques as a complementary probe to conventional transport experiments in the study of superconductor heterostructures, and enables a deeper understanding of induced unconventional superconductivity in these systems.

More broadly, our technique is applicable to a wide variety of mesoscopic superconducting systems, as realized in other proximity-coupled systems such as superconductor/semiconductor heterostructures \cite{higginbotham-prl}, interfacial superconductivity in oxide heterostructures, and two-dimensional superconducting materials such as graphene heterostructures or transition metal dichalcogenides \cite{cboettcher2022}. Consequently, our technique can be leveraged as a novel means to directly probe (potentially unconventional) superconductivity in a wide array of exotic low-dimensional systems that have thus far been challenging to probe via conventional techniques. \\

\begin{acknowledgments}
\textit{Acknowledgements:} The authors thank Eugene Demler, Bertrand Halperin, Jonathan Curtis, and Leonid Glazman for fruitful discussions relating to this work. 
The experimental work is supported by the Quantum Science Center (QSC), a National Quantum Information Science Research Center of  the  U.S.  Department  of  Energy  (DOE). Device fabrication was performed at the Center for Nanoscale Systems at Harvard, supported in part by an NSF NNIN
award ECS-00335765. N.R.P. and M.E.W. are supported by the Department of Defense through the NDSEG fellowship program. 
A.G. and V.M.G. are supported by the National Science Foundation under Grant No. DMR-2037158, the U.S. Army Research Office under Contract No. W911NF1310172, and the Simons Foundation.  A.Y. is partly supported by the Gordon and Betty Moore Foundation through Grant GBMF 9468 and by the National Science Foundation under Grant No. DMR-1708688. 
\end{acknowledgments}

\begin{appendix}
    \section{Technical details of the calculation of the superfluid density\label{sec:Technical-details-of}}

We now discuss the technical details of calculating the superfluid
density in the simplified phenomenological model. The clean-limit of this theory agrees with the experimental results presented in the main text, while the dirty limit of the theory leads to different behavior. This limit, studied using the Usadel equation, will be presented elsewhere.

We begin by discussing the generation of the nodal $p$-wave condensate component inside
the ferromagnet. We note that while a detailed calculation can be
found in \cite{takei_SF}, here we only study a simplified model described
by the 2x2 Bogolyubov-de-Gennes Hamiltonian:

\begin{align*}
 & H_{\text{BdG}}=\\
 & \left(\begin{array}{cc}
\xi_{{\bf k}}^{\left(\text{s}\right)}\hat{\tau}_{3}+\Delta\hat{\tau}_{2}\hat{\sigma}_{2} & t\hat{\tau}_{3}+t_{\text{soc}}\left({\bf k}_{x}\hat{\sigma}_{2}\hat{\tau}_{3}-{\bf k}_{y}\hat{\sigma}_{1}\right)\\
t\hat{\tau}_{3}+t_{\text{soc}}\left({\bf k}_{x}\hat{\sigma}_{2}\hat{\tau}_{3}-{\bf k}_{y}\hat{\sigma}_{1}\right) & \xi_{{\bf k}}^{\left(\text{f}\right)}\hat{\tau}_{3}+h_{x}\hat{\tau}_{3}\hat{\sigma}_{1}
\end{array}\right),\\
\end{align*}
where $\xi_{{\bf k}}^{(\text{s}/\text{f})}$ are the electronic dispersions,
$h_{x}$ is the Zeeman field value, $\Delta$ is the bulk gap inside
the superconductor, $t$ is the regular tunneling and $t_{\text{soc}}$
is the tunneling with spin-orbit interaction, the Nambu and spin Pauli
matrices are denoted as $\hat{\tau}_{i}$ and $\hat{\sigma}_{i}$.
Using projector formalism, we now integrate the superconductor and
the minority-spin component degree of freedom. The induced imaginary-frequency
self-energy reads:

\begin{align*}
\hat{\Sigma}_{{\bf k}}\left(i\epsilon_{n}\right) & =-{\cal P}H_{\text{BdG}}\left(i\epsilon_{n}-QH_{\text{BdG}}Q\right)^{-1}H_{\text{BdG}}{\cal P}
\end{align*}
where ${\cal P}$ is the projector onto majority-spin component, ${\cal Q}=\mathbb{I}-{\cal P}$
and $\epsilon_{n}=(2n+1)\pi/\beta$. The resulting expression can
be found analytically by expanding up to the second order in the tunneling
but it is still too cumbersome to be reproduced here. Importantly,
depending on the Fermi surface geometry and scattering properties,
there are two kinds of terms: induced spin-orbit interaction $\sim\frac{2k_{y}tt_{\text{soc}}\hat{\tau}_{0}}{\Delta^{2}+\delta E^{2}}\delta E$
and the nodal $p$-wave triplet component $\sim\frac{2k_{x}tt_{\text{soc}}\Delta}{\Delta^{2}+\delta E^{2}}\hat{\tau}_{2}$,
where $\delta E$ is the difference of Fermi energies of the majority-spin
and the superconductor. In the following, we focus only on the triplet
component assuming $\hat{\Sigma}_{{\bf k}}\approx k_{x}\Delta_{t}\hat{\tau}_{2}$
with $\Delta_{t}$ being a free parameter.

\subsection{Self-energy and vertex corrections}
We now consider the disorder averaging and vertex corrections to the
superfluid density. Within the self-consistent $T$-matrix approximation,
the self-energy due to disorder scattering reads \cite{mahan_txtbk}:

\[
\hat{\Sigma}\left(i\epsilon_{n}\right)=n_{i}\hat{T}\left(i\epsilon_{n}\right),
\]
where $n_{i}$ is the impurity concentration. The $\hat{T}$-matrix
is given by the sum of ladder diagrams and is equal to:

\[
\hat{T}\left(i\epsilon_{n}\right)=v_{0}\left(1-v_{0}\left\langle \hat{G}_{{\bf k}}\left(i\epsilon_{n}\right)\right\rangle \right)^{-1},
\]
where $\left\langle \ldots\right\rangle =L^{-2}\sum_{k}$ and $v_{0}$
is the disorder scattering strength and $G_{{\bf k}}$ being full
Green's function. In the main text, we take the limit $v_{0}\rightarrow\infty$
and denote the scattering rate as $\tau^{-1}=n_{i}/\nu_{0}$, where $\nu_{0}$ is the electronic density of states. In this
case, the $T$-matrix is $\hat{T}\left(i\epsilon_{n}\right)=T\left(i\epsilon_{n}\right)\hat{\tau}_{3}$,
and the remaining equation for $T(i\epsilon_{n})$ can be solved self-consistently
to find the self-energy. 

We also need to consider the proper vertex corrections to compute
the Meissner response.Within the self-consistent $T$-matrix approximation
\cite{mahan_txtbk} the correction to the current vertex $\Gamma_{\mu}$
are given by the Bethe-Salpeter equation:

\onecolumngrid

\[
\Gamma_{\mu}\left(i\epsilon_{n},i\epsilon_{n}+i\Omega_{m},{\bf k}\right)=\gamma_{\mu}\left({\bf k}\right)+n_{i}\int\frac{d^{2}{\bf k}'}{(2\pi)^{2}}T\left(i\epsilon_{n}\right)T\left(i\epsilon_{n}+i\Omega_{m}\right)\tau_{3}\hat{G}_{{\bf k}'}\left(i\epsilon_{n}+i\Omega_{m}\right)\Gamma_{\mu}(i\epsilon_{n},i\epsilon_{n}+i\Omega_{m},{\bf k})\hat{G}_{{\bf k'}}(i\epsilon_{n})\tau_{3}.
\]

\twocolumngrid

Following \cite{mahan_txtbk}, we first analytically continue this
equation to real frequencies and then solve it numerically.

\section{Device fabrication}
The devices studied in this work are fabricated by first thermally evaporating gold bond pads and alignment marks onto a high-resistivity silicon chip. Next, the chip is dipped in hydroflouric (HF) acid and a 25 nm thick Nb film is immediately sputtered onto the cleaned chip. The resonator structure is defined using electron-beam lithography and the unwanted Nb is removed via reactive ion etching with CF$_4$. To fabricate the S/F hybrid resonators, the S/F bilayer region at the end of the resonator is defined in another electron-beam lithography step. The exposed Nb region is cleaned in HF to ensure a transparent interface, after which a 30 nm thick permalloy film is immediately thermally evaporated.

\section{Experimental setup}
\begin{figure}[t]
	\includegraphics[width= 3.5 in]{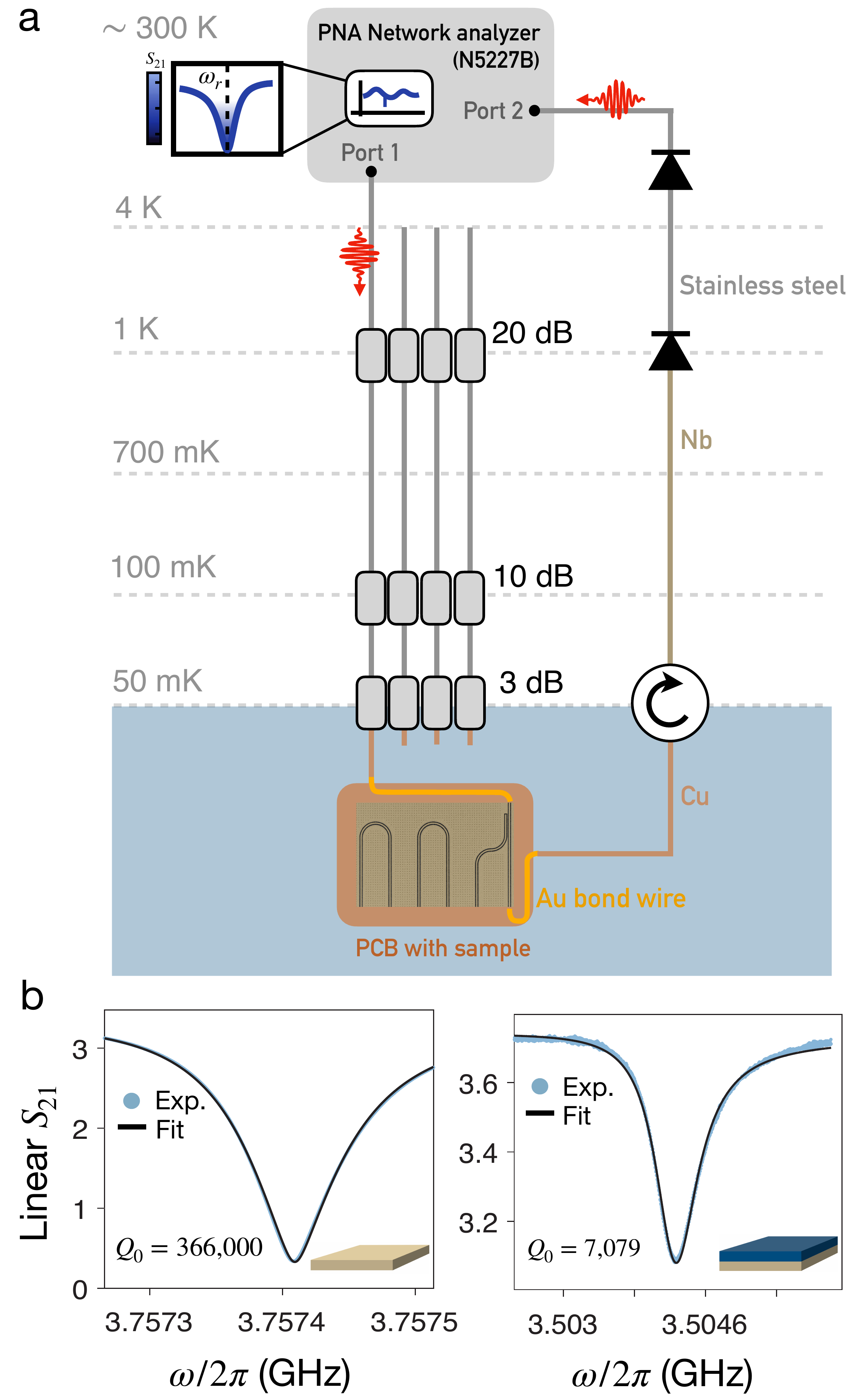} 
	\caption{\textbf{Experimental setup}. \textbf{a.} Schematic wiring diagram for the microwave measurement setup. All lines are coaxial cables, with the materials for each segment indicated in the figure. Grey boxes represent attenuators thermally anchored to each plate of the dilution refrigerator. \textbf{b.} Traces of the microwave transmission $S_{21}$ for bare Nb resonators and hybrid resonators terminated with an S/F bilayer, respectively. The fit used to extract the resonance frequency is shown along with the raw data, along with the quality factor estimated from the fit.}
 \label{exp-fig}
\end{figure}

The experiments described in this work are performed in a dilution refrigerator (Oxford Instruments Kelvinox MX50) with a base temperature of 55 mK equipped with a three-axis vector magnet. Microwave signals generated by a vector network analyzer (Keysight PNA Microwave Network Analyzer N5227B) are sent down a stainless steel coaxial line which is thermally anchored via attenuators to each plate of the cryostat as illustrated in Fig. \ref{exp-fig}a. The sample is mounted on the mixing chamber in a copper sample holder designed by IBM Research, and is connected to the measurement  circuit via non-superconducting gold wirebonds. The signal transmitted across the device is routed through a circulator (Quinstar QCY-G0400801) via Nb superconducting coaxial lines to a cold amplifier (Low Noise Factory LNF-LNC03-14SA) on the 4K plate. The amplified signal then leaves the cryostat via stainless steel coaxial lines and is further amplified at room temperature (MITEQ LNA-40-04000800-07-10P) before being read out into one of the ports of the network analyzer. 

The microwave transmission $S_{21}$ is recorded and fit to the standard form for a hanging resonator \cite{hanging-res} to extract the resonance frequency. Representative traces of $S_{21}$ versus frequency featuring the resonator mode for devices terminated both with and without the S/F bilayer are shown in Fig. \ref{exp-fig}b, along with the corresponding fits used to extract the resonance frequency.

\section{Measuring the superconducting gap of aluminum}
\begin{figure}[t]
	\includegraphics[width= 2.7 in]{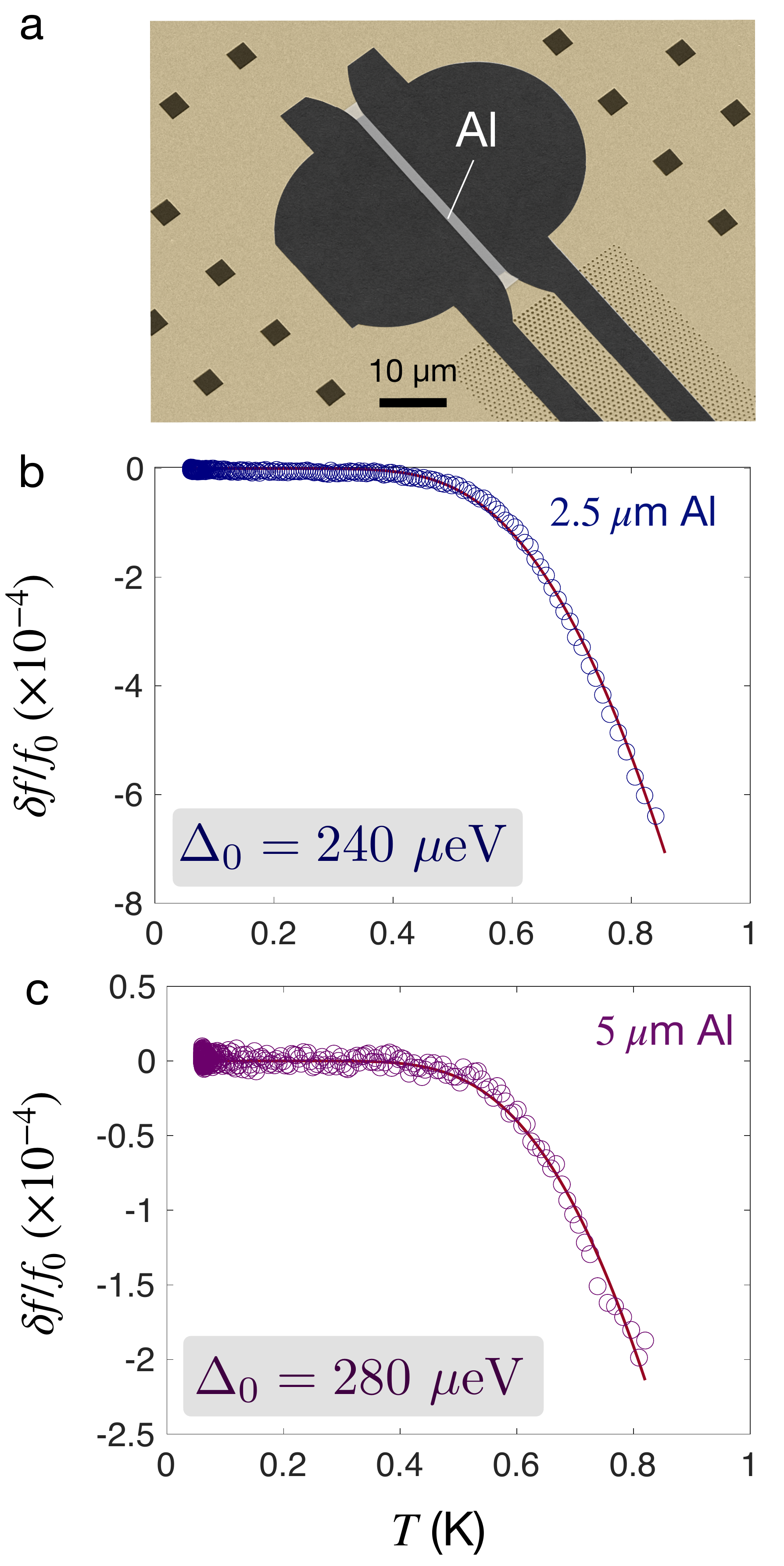} 
	\caption{\textbf{Aluminum hybrid resonators}. \textbf{a.} False-colored scanning electron micrograph of the Al strip terminating a Nb resonator otherwise identical to that used in the S/F bilayer devices, as described at length in the main text. \textbf{b.} Temperature-dependent resonance frequency of a device with a 2.5 $\mu$m wide Al strip. The data is fit to Eq. (\ref{bcs-sf}), which yields a value of $\Delta_0 = 240$ $\mu$eV for the superconducting gap. \textbf{c.} Temperature-dependent resonance frequency and BCS fit for a device with a 5 $\mu$m wide Al strip.}
 \label{al-fig}
\end{figure}

As a proof of concept for our kinetic inductance measurement technique, we studied a device in which the Nb resonator is shorted to ground through a small Al strip, rather than an S/F bilayer. The termination of the resonator is shown in Fig. \ref{al-fig}a: the center conductor of the Nb resonator is ``cut'' and replaced with a 40 $\mu$m long, 20 nm thick Al film with a width of either 2.5 $\mu$m or 5 $\mu$m. As discussed in the main text, the resonator design localizes most of the current to the Al strip, making the resonator response particularly sensitive to the Al region. Moreover, since the critical temperature and superconducting gap of Al are much lower than that of Nb, the temperature dependence of the resonator frequency will be almost exclusively due to the temperature dependence of the superfluid density in the Al strip. In Fig. \ref{al-fig}b,c we show the temperature dependence of the resonance frequency for the two resonators with different widths of the Al strips. In both cases, the curves are activated with temperature, as one would expect for a fully-gapped conventional superconductor. Moreover, we may fit these data to the low-temperature limit of the standard BCS form for the superfluid density \cite{Prozorov},
   \begin{equation} \label{bcs-sf}
       \frac{\delta f}{f_0} = A \sqrt{\frac{2\pi \Delta_0}{T}} \; \e^{-\Delta_0/T} \, .
   \end{equation}
The amplitude of the frequency shift, $A$, and the zero-temperature superconducting gap $\Delta_0$ are treated as fit parameters. The fits to each curve are superimposed on the data in Fig. \ref{al-fig}b,c, and yield values of the gap $\Delta_0 = 240$ $\mu$eV (280 $\mu$eV) for the resonator with a 2.5 $\mu$m (5 $\mu$m) Al strip. These values are consistent with direct measurements of the superconducting gap of Al thin films, which quantitatively validates our measurement technique and analysis procedure.

\section{Magnon-photon coupling}
As described in the main text, by tuning an external field such that the Kittel mode frequency, $\omega_m = \gamma\sqrt{\mu_0^2H_{\parallel}( H_{\parallel}+M_s)}$, coincides with that of the resonator, $\omega_r$, avoided crossings are observed symmetrically around zero for the first and third harmonic of the resonator.  When the magnons couple to the photons, the hybrid mode is highly broadened due to magnon damping. We obtain the magnon-photon coupling strength, $g$, from modeling the two bands, using the following equation for the transmission spectrum \cite{blais-rmp,nb-py-prl1,nb-py-prl2}
\begin{equation}
    S_{21}(\omega,H_{\parallel})=\frac{\kappa_{r,ext}}{i(\omega-\omega_r)-\kappa_r+\frac{g^2}{i[\omega-\omega_m(H_\parallel)]-\kappa_m/2}}
\end{equation}
where $\kappa_r$ and $\kappa_{r,ext}$ are the resonator internal and external loss rate respectively and $\kappa_m$ is the magnon damping rate. From this we determine the total magnon-photon coupling to be $g_{\mathrm{1st}}=120$ MHz for the first harmonic, consistent with the coupling extracted from anticrossing of the third harmonic of the resonator, $g_{\mathrm{3rd}}=100$ MHz. We note that the extracted saturation magnetization of the first mode, $M_s^{\mathrm{1st}}$ is more than twice the value that we obtain from the third harmonic, $\mu_0 M^{\mathrm{3rd}}_s=1.38$ T, and what has previously been reported in the literature \cite{nb-py-prl1,nb-py-prl2}. We speculate that trapped fields could lead to a seemingly larger saturation magnetization.

\section{Alignment procedure}
\begin{figure}[t]
	\includegraphics[width= 3 in]{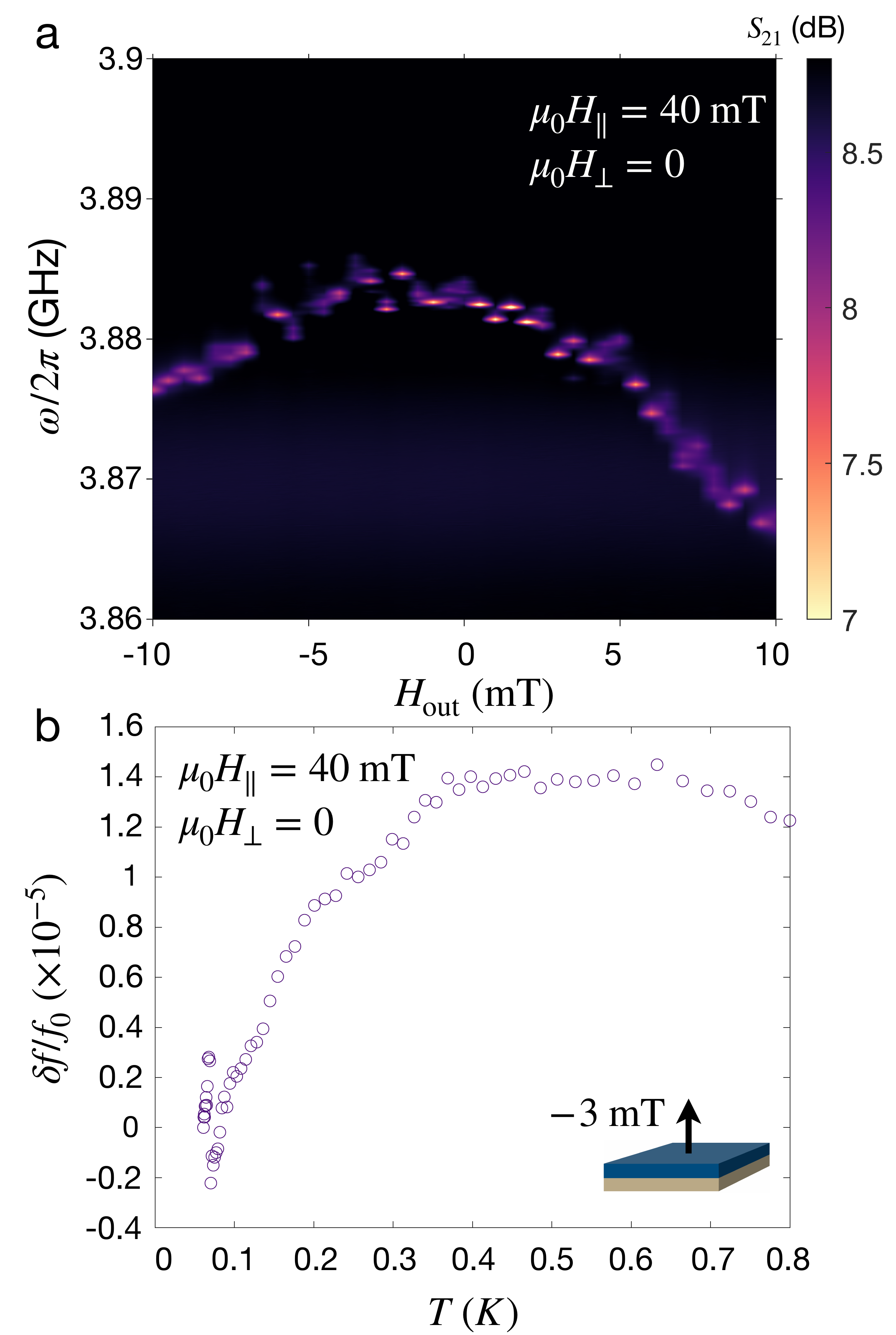} 
	\caption{\textbf{Field alignment procedure}. \textbf{a.} Resonance frequency of a S/F hybrid resonator as a function of the applied out-of-plane magnetic field. The maximum frequency of the resonator corresponds to the effective zero-field. \textbf{b.} Example of a temperature scan of the resonance frequency when the field alignment procedure is not followed and the temperature-dependence exhibits non-systematic behavior. }
 \label{parabola-fig}
\end{figure}

The measurements described in the main text are performed in an in-plane magnetic field, although in reality sample misalignment inevitably leads to small out-of-plane field components. To eliminate the effects of these small unwanted out-of-plane fields, we employ the three-axis vector magnet in our cryostat to compensate for the out-of-plane field and ensure that the magnetic field experienced by the sample is entirely in-plane. To do so, we use the resonator frequency as a sensitive measure of the field experienced by the sample. The resonance frequency of the superconducting resonator decreases with applied out-of-plane fields in an approximately parabolic fashion for small fields due to field-induced pair breaking and the associated decrease in superfluid density (or, equivalently, increase in kinetic inductance), as shown in Fig. \ref{parabola-fig}a. The maximum of this parabola indicates the ``effective'' zero-field where the out-of-plane field experienced by the sample vanishes. 

To align the magnetic field, we begin by setting the nominal in-plane field to its desired value. At this fixed in-plane field, we sweep the out-of-plane field and determine the applied out-of-plane field corresponding to the effective zero-field as described above. We then re-trace the out-of-plane field history to avoid any hysteretic effects and set the out-of-plane field to it's effective-zero-field value (i.e. the applied field corresponding to the maximum of the parabola). With this field configuration fixed, we then proceed with our temperature-dependent scans. Similar techniques for field alignment have been employed in previous studies of superconducting resonators \cite{nlme-cpw}. 

If this alignment procedure is not followed, temperature-dependent resonance frequency traces often feature artifacts due to trapped vortices. An example of one such effect, a pronounced ``downturn'' in the resonance frequency as the temperature is lowered, is shown in Fig. \ref{parabola-fig}b. These artifacts are strongly history-dependent and non-systematic. In contrast, when the alignment procedure described above is followed (as is the case for all data presented throughout the main text), the temperature-dependent traces are free of these artifacts and are highly reproducible.

\section{Niobium resonators in magnetic fields}
\begin{figure*}[t]
	\includegraphics[width= 6.5 in]{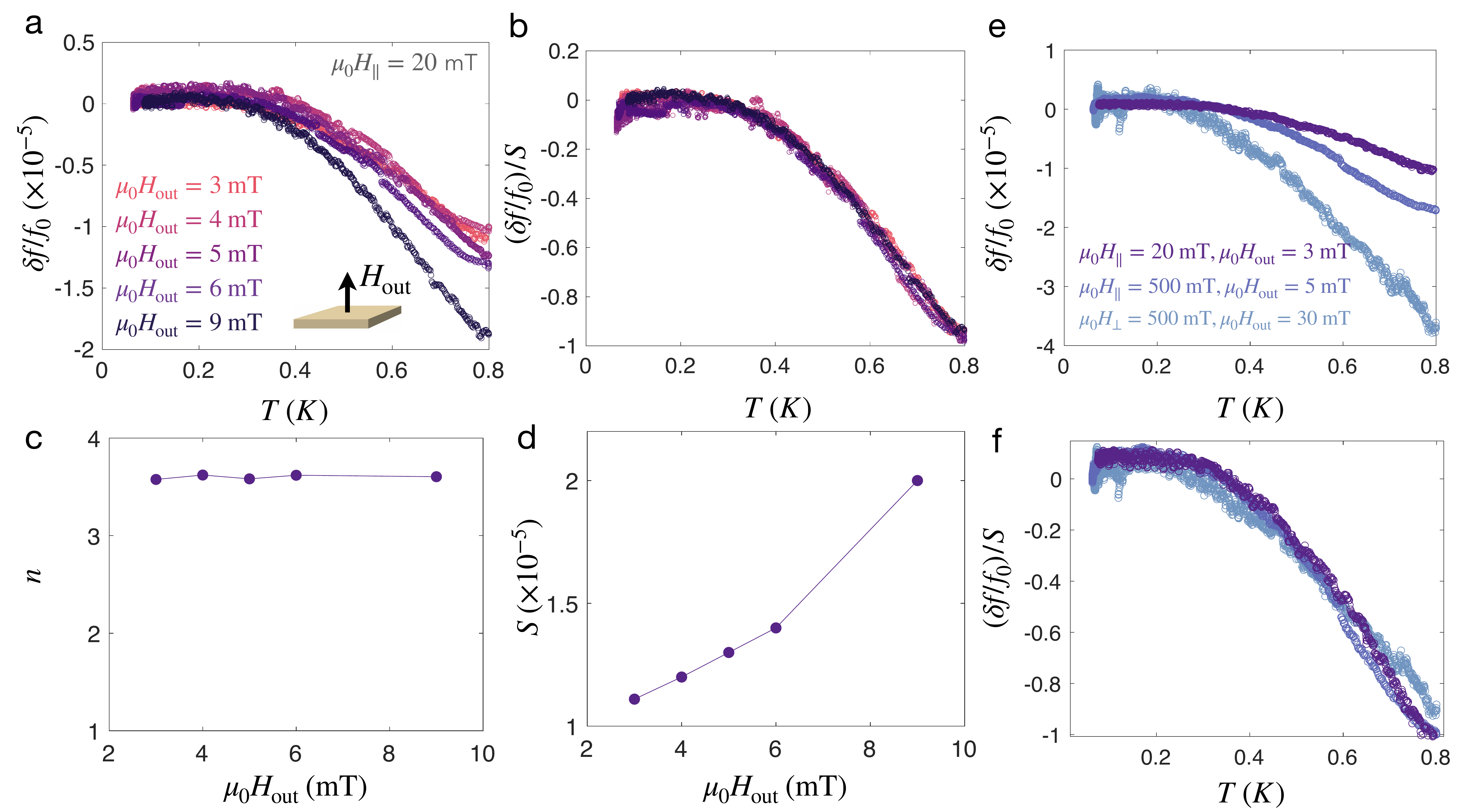} 
	\caption{\textbf{Niobium resonators in magnetic fields}. \textbf{a.} Temperature-dependent resonance frequency of a Nb resonator subject to an in-plane magnetic field $\mu_0 H_\parallel = 20 mT$ for different values of the out-of-plane field. \textbf{b.} Data in panel \textbf{a} normalized to the integrated frequency shift $S$. \textbf{c.,d.} Extracted temperature scaling exponent $n$ and frequency shift $S$ for the data in panel \textbf{a}. \textbf{e} Temperature-dependent resonance frequency for Nb resonators subjected to in-plane magnetic fields in both the $H_\parallel$ and $H_\perp$ directions. \textbf{e.} Data from panel \textbf{d} normalized to the total frequency shift $S$. }
 \label{field-fig}
\end{figure*}

In this section we discuss the phenomenology of bare Nb resonators (i.e. without S/F bilayers) subject to magnetic fields, and contrast their behavior to that of the hybrid S/F resonators studied in the main text. In Fig. \ref{field-fig}a we show temperature-dependent traces of the resonant frequency for a Nb resonator subject to an in-plane field of $\mu_0 H_\parallel = 20$ mT for varying values of the out-of-plane magnetic field. To compare to the results in the main text, we fit the response to a power law $\delta f/f_0 = AT^n$, and take the phenomenological approach of considering exponents $n \approx 4$ equivalent to an activated temperature dependence. We also introduce the total frequency shift $S = [f(55 \;\text{mK}) - f(800 \;\text{mK})]/f(55 \;\text{mK})$ which quantifies the overall size of the frequency shift with temperature in each run. We plot the extracted $S$ and $n$ as a function of the out-of-plane field in Fig. \ref{field-fig}c,d, where we see that the temperature-scaling exponent is unchanged by the out-of-plane field. In contrast, the net frequency shift $S$ increases monotonically with the out-of-plane field, presumably due to the reduction of the superfluid density and commensurate decrease of $f_0$.  To emphasize the insensitivity of the temperature scaling to magnetic fields, in Fig. \ref{field-fig}b we normalize the frequency shifts to $S$, and see that the curves for each out-of-plane field collapse onto one another. Thus, the sole effect of the out-of-plane magnetic field is to rescale the total size of the frequency shift, and does not affect the temperature scaling in any way.

Further, in Fig. \ref{field-fig}e we compare temperature-dependent traces in in-plane fields along both orthogonal directions $H_\parallel$ and $H_\perp$ studied in the main text. We note that in this particular cooldown, the misalignment in the $H_\perp$ direction was somewhat large, with $H_{\text{out}} \approx 20$ mT corresponding to the effective zero field. Again, independent of the out-of-plane magnetic field, all curves display an activated temperature dependence. Normalizing by $S$ as before, we find that the temperature-dependent traces for both in-plane field directions collapse onto one another. That is, the temperature dependent response of bare Nb resonators is activated irrespective of applied magnetic fields and isotropic with respect to the orientation of in-plane fields.

\end{appendix}

\bibliography{refs}

\begin{thebibliography}{49}%
\makeatletter
\providecommand \@ifxundefined [1]{%
 \@ifx{#1\undefined}
}%
\providecommand \@ifnum [1]{%
 \ifnum #1\expandafter \@firstoftwo
 \else \expandafter \@secondoftwo
 \fi
}%
\providecommand \@ifx [1]{%
 \ifx #1\expandafter \@firstoftwo
 \else \expandafter \@secondoftwo
 \fi
}%
\providecommand \natexlab [1]{#1}%
\providecommand \enquote  [1]{``#1''}%
\providecommand \bibnamefont  [1]{#1}%
\providecommand \bibfnamefont [1]{#1}%
\providecommand \citenamefont [1]{#1}%
\providecommand \href@noop [0]{\@secondoftwo}%
\providecommand \href [0]{\begingroup \@sanitize@url \@href}%
\providecommand \@href[1]{\@@startlink{#1}\@@href}%
\providecommand \@@href[1]{\endgroup#1\@@endlink}%
\providecommand \@sanitize@url [0]{\catcode `\\12\catcode `\$12\catcode
  `\&12\catcode `\#12\catcode `\^12\catcode `\_12\catcode `\%12\relax}%
\providecommand \@@startlink[1]{}%
\providecommand \@@endlink[0]{}%
\providecommand \url  [0]{\begingroup\@sanitize@url \@url }%
\providecommand \@url [1]{\endgroup\@href {#1}{\urlprefix }}%
\providecommand \urlprefix  [0]{URL }%
\providecommand \Eprint [0]{\href }%
\providecommand \doibase [0]{http://dx.doi.org/}%
\providecommand \selectlanguage [0]{\@gobble}%
\providecommand \bibinfo  [0]{\@secondoftwo}%
\providecommand \bibfield  [0]{\@secondoftwo}%
\providecommand \translation [1]{[#1]}%
\providecommand \BibitemOpen [0]{}%
\providecommand \bibitemStop [0]{}%
\providecommand \bibitemNoStop [0]{.\EOS\space}%
\providecommand \EOS [0]{\spacefactor3000\relax}%
\providecommand \BibitemShut  [1]{\csname bibitem#1\endcsname}%
\let\auto@bib@innerbib\@empty
\bibitem [{\citenamefont {Sato}\ and\ \citenamefont
  {Ando}(2017)}]{Sato_review}%
  \BibitemOpen
  \bibfield  {author} {\bibinfo {author} {\bibfnamefont {M.}~\bibnamefont
  {Sato}}\ and\ \bibinfo {author} {\bibfnamefont {Y.}~\bibnamefont {Ando}},\
  }\href {\doibase 10.1088/1361-6633/aa6ac7} {\bibfield  {journal} {\bibinfo
  {journal} {Reports on Progress in Physics}\ }\textbf {\bibinfo {volume}
  {80}},\ \bibinfo {pages} {076501} (\bibinfo {year} {2017})}\BibitemShut
  {NoStop}%
\bibitem [{\citenamefont {Linder}\ and\ \citenamefont
  {Robinson}(2015)}]{sc-spintronics}%
  \BibitemOpen
  \bibfield  {author} {\bibinfo {author} {\bibfnamefont {J.}~\bibnamefont
  {Linder}}\ and\ \bibinfo {author} {\bibfnamefont {J.~W.~A.}\ \bibnamefont
  {Robinson}},\ }\href {\doibase 10.1038/nphys3242} {\bibfield  {journal}
  {\bibinfo  {journal} {Nature Physics}\ }\textbf {\bibinfo {volume} {11}},\
  \bibinfo {pages} {307} (\bibinfo {year} {2015})}\BibitemShut {NoStop}%
\bibitem [{\citenamefont {Nayak}\ \emph {et~al.}(2008)\citenamefont {Nayak},
  \citenamefont {Simon}, \citenamefont {Stern}, \citenamefont {Freedman},\ and\
  \citenamefont {Das~Sarma}}]{sds-rmp}%
  \BibitemOpen
  \bibfield  {author} {\bibinfo {author} {\bibfnamefont {C.}~\bibnamefont
  {Nayak}}, \bibinfo {author} {\bibfnamefont {S.~H.}\ \bibnamefont {Simon}},
  \bibinfo {author} {\bibfnamefont {A.}~\bibnamefont {Stern}}, \bibinfo
  {author} {\bibfnamefont {M.}~\bibnamefont {Freedman}}, \ and\ \bibinfo
  {author} {\bibfnamefont {S.}~\bibnamefont {Das~Sarma}},\ }\href {\doibase
  10.1103/RevModPhys.80.1083} {\bibfield  {journal} {\bibinfo  {journal} {Rev.
  Mod. Phys.}\ }\textbf {\bibinfo {volume} {80}},\ \bibinfo {pages} {1083}
  (\bibinfo {year} {2008})}\BibitemShut {NoStop}%
\bibitem [{\citenamefont {Volovik}(2009)}]{volovik-3heb}%
  \BibitemOpen
  \bibfield  {author} {\bibinfo {author} {\bibfnamefont {G.~E.}\ \bibnamefont
  {Volovik}},\ }\href {\doibase 10.1134/S0021364009170172} {\bibfield
  {journal} {\bibinfo  {journal} {JETP Letters}\ }\textbf {\bibinfo {volume}
  {90}},\ \bibinfo {pages} {398} (\bibinfo {year} {2009})}\BibitemShut
  {NoStop}%
\bibitem [{\citenamefont {Sauls}(1994)}]{jim-upt3}%
  \BibitemOpen
  \bibfield  {author} {\bibinfo {author} {\bibfnamefont {J.}~\bibnamefont
  {Sauls}},\ }\href {\doibase 10.1080/00018739400101475} {\bibfield  {journal}
  {\bibinfo  {journal} {Advances in Physics}\ }\textbf {\bibinfo {volume}
  {43}},\ \bibinfo {pages} {113} (\bibinfo {year} {1994})}\BibitemShut
  {NoStop}%
\bibitem [{\citenamefont {Avers}\ \emph {et~al.}(2020)\citenamefont {Avers},
  \citenamefont {Gannon}, \citenamefont {Kuhn}, \citenamefont {Halperin},
  \citenamefont {Sauls}, \citenamefont {DeBeer-Schmitt}, \citenamefont
  {Dewhurst}, \citenamefont {Gavilano}, \citenamefont {Nagy}, \citenamefont
  {Gasser},\ and\ \citenamefont {Eskildsen}}]{ins-upt3}%
  \BibitemOpen
  \bibfield  {author} {\bibinfo {author} {\bibfnamefont {K.~E.}\ \bibnamefont
  {Avers}}, \bibinfo {author} {\bibfnamefont {W.~J.}\ \bibnamefont {Gannon}},
  \bibinfo {author} {\bibfnamefont {S.~J.}\ \bibnamefont {Kuhn}}, \bibinfo
  {author} {\bibfnamefont {W.~P.}\ \bibnamefont {Halperin}}, \bibinfo {author}
  {\bibfnamefont {J.~A.}\ \bibnamefont {Sauls}}, \bibinfo {author}
  {\bibfnamefont {L.}~\bibnamefont {DeBeer-Schmitt}}, \bibinfo {author}
  {\bibfnamefont {C.~D.}\ \bibnamefont {Dewhurst}}, \bibinfo {author}
  {\bibfnamefont {J.}~\bibnamefont {Gavilano}}, \bibinfo {author}
  {\bibfnamefont {G.}~\bibnamefont {Nagy}}, \bibinfo {author} {\bibfnamefont
  {U.}~\bibnamefont {Gasser}}, \ and\ \bibinfo {author} {\bibfnamefont {M.~R.}\
  \bibnamefont {Eskildsen}},\ }\href {\doibase 10.1038/s41567-020-0822-z}
  {\bibfield  {journal} {\bibinfo  {journal} {Nature Physics}\ }\textbf
  {\bibinfo {volume} {16}},\ \bibinfo {pages} {531} (\bibinfo {year}
  {2020})}\BibitemShut {NoStop}%
\bibitem [{\citenamefont {Joynt}\ and\ \citenamefont
  {Taillefer}(2002)}]{upt3-rmp}%
  \BibitemOpen
  \bibfield  {author} {\bibinfo {author} {\bibfnamefont {R.}~\bibnamefont
  {Joynt}}\ and\ \bibinfo {author} {\bibfnamefont {L.}~\bibnamefont
  {Taillefer}},\ }\href {\doibase 10.1103/RevModPhys.74.235} {\bibfield
  {journal} {\bibinfo  {journal} {Rev. Mod. Phys.}\ }\textbf {\bibinfo {volume}
  {74}},\ \bibinfo {pages} {235} (\bibinfo {year} {2002})}\BibitemShut
  {NoStop}%
\bibitem [{\citenamefont {Aoki}\ and\ \citenamefont {Flouquet}(2014)}]{u-sc}%
  \BibitemOpen
  \bibfield  {author} {\bibinfo {author} {\bibfnamefont {D.}~\bibnamefont
  {Aoki}}\ and\ \bibinfo {author} {\bibfnamefont {J.}~\bibnamefont
  {Flouquet}},\ }\href {\doibase 10.7566/JPSJ.83.061011} {\bibfield  {journal}
  {\bibinfo  {journal} {Journal of the Physical Society of Japan}\ }\textbf
  {\bibinfo {volume} {83}},\ \bibinfo {pages} {061011} (\bibinfo {year}
  {2014})}\BibitemShut {NoStop}%
\bibitem [{\citenamefont {Ran}\ \emph {et~al.}(2019)\citenamefont {Ran},
  \citenamefont {Eckberg}, \citenamefont {Ding}, \citenamefont {Furukawa},
  \citenamefont {Metz}, \citenamefont {Saha}, \citenamefont {Liu},
  \citenamefont {Zic}, \citenamefont {Kim}, \citenamefont {Paglione},\ and\
  \citenamefont {Butch}}]{ute2}%
  \BibitemOpen
  \bibfield  {author} {\bibinfo {author} {\bibfnamefont {S.}~\bibnamefont
  {Ran}}, \bibinfo {author} {\bibfnamefont {C.}~\bibnamefont {Eckberg}},
  \bibinfo {author} {\bibfnamefont {Q.-P.}\ \bibnamefont {Ding}}, \bibinfo
  {author} {\bibfnamefont {Y.}~\bibnamefont {Furukawa}}, \bibinfo {author}
  {\bibfnamefont {T.}~\bibnamefont {Metz}}, \bibinfo {author} {\bibfnamefont
  {S.~R.}\ \bibnamefont {Saha}}, \bibinfo {author} {\bibfnamefont {I.-L.}\
  \bibnamefont {Liu}}, \bibinfo {author} {\bibfnamefont {M.}~\bibnamefont
  {Zic}}, \bibinfo {author} {\bibfnamefont {H.}~\bibnamefont {Kim}}, \bibinfo
  {author} {\bibfnamefont {J.}~\bibnamefont {Paglione}}, \ and\ \bibinfo
  {author} {\bibfnamefont {N.~P.}\ \bibnamefont {Butch}},\ }\href {\doibase
  10.1126/science.aav8645} {\bibfield  {journal} {\bibinfo  {journal}
  {Science}\ }\textbf {\bibinfo {volume} {365}},\ \bibinfo {pages} {684}
  (\bibinfo {year} {2019})}\BibitemShut {NoStop}%
\bibitem [{\citenamefont {Matano}\ \emph {et~al.}(2016)\citenamefont {Matano},
  \citenamefont {Kriener}, \citenamefont {Segawa}, \citenamefont {Ando},\ and\
  \citenamefont {Zheng}}]{cubise}%
  \BibitemOpen
  \bibfield  {author} {\bibinfo {author} {\bibfnamefont {K.}~\bibnamefont
  {Matano}}, \bibinfo {author} {\bibfnamefont {M.}~\bibnamefont {Kriener}},
  \bibinfo {author} {\bibfnamefont {K.}~\bibnamefont {Segawa}}, \bibinfo
  {author} {\bibfnamefont {Y.}~\bibnamefont {Ando}}, \ and\ \bibinfo {author}
  {\bibfnamefont {G.-q.}\ \bibnamefont {Zheng}},\ }\href {\doibase
  10.1038/nphys3781} {\bibfield  {journal} {\bibinfo  {journal} {Nature
  Physics}\ }\textbf {\bibinfo {volume} {12}},\ \bibinfo {pages} {852}
  (\bibinfo {year} {2016})}\BibitemShut {NoStop}%
\bibitem [{\citenamefont {Zhou}\ \emph {et~al.}(2022)\citenamefont {Zhou},
  \citenamefont {Holleis}, \citenamefont {Saito}, \citenamefont {Cohen},
  \citenamefont {Huynh}, \citenamefont {Patterson}, \citenamefont {Yang},
  \citenamefont {Taniguchi}, \citenamefont {Watanabe},\ and\ \citenamefont
  {Young}}]{bbg-sc}%
  \BibitemOpen
  \bibfield  {author} {\bibinfo {author} {\bibfnamefont {H.}~\bibnamefont
  {Zhou}}, \bibinfo {author} {\bibfnamefont {L.}~\bibnamefont {Holleis}},
  \bibinfo {author} {\bibfnamefont {Y.}~\bibnamefont {Saito}}, \bibinfo
  {author} {\bibfnamefont {L.}~\bibnamefont {Cohen}}, \bibinfo {author}
  {\bibfnamefont {W.}~\bibnamefont {Huynh}}, \bibinfo {author} {\bibfnamefont
  {C.~L.}\ \bibnamefont {Patterson}}, \bibinfo {author} {\bibfnamefont
  {F.}~\bibnamefont {Yang}}, \bibinfo {author} {\bibfnamefont {T.}~\bibnamefont
  {Taniguchi}}, \bibinfo {author} {\bibfnamefont {K.}~\bibnamefont {Watanabe}},
  \ and\ \bibinfo {author} {\bibfnamefont {A.~F.}\ \bibnamefont {Young}},\
  }\href {\doibase 10.1126/science.abm8386} {\bibfield  {journal} {\bibinfo
  {journal} {Science}\ }\textbf {\bibinfo {volume} {375}},\ \bibinfo {pages}
  {774} (\bibinfo {year} {2022})}\BibitemShut {NoStop}%
\bibitem [{\citenamefont {Lutchyn}\ \emph {et~al.}(2010)\citenamefont
  {Lutchyn}, \citenamefont {Sau},\ and\ \citenamefont
  {Das~Sarma}}]{lsd-majorana}%
  \BibitemOpen
  \bibfield  {author} {\bibinfo {author} {\bibfnamefont {R.~M.}\ \bibnamefont
  {Lutchyn}}, \bibinfo {author} {\bibfnamefont {J.~D.}\ \bibnamefont {Sau}}, \
  and\ \bibinfo {author} {\bibfnamefont {S.}~\bibnamefont {Das~Sarma}},\ }\href
  {\doibase 10.1103/PhysRevLett.105.077001} {\bibfield  {journal} {\bibinfo
  {journal} {Phys. Rev. Lett.}\ }\textbf {\bibinfo {volume} {105}},\ \bibinfo
  {pages} {077001} (\bibinfo {year} {2010})}\BibitemShut {NoStop}%
\bibitem [{\citenamefont {Alicea}(2012)}]{Alicea_majorana}%
  \BibitemOpen
  \bibfield  {author} {\bibinfo {author} {\bibfnamefont {J.}~\bibnamefont
  {Alicea}},\ }\href {\doibase 10.1088/0034-4885/75/7/076501} {\bibfield
  {journal} {\bibinfo  {journal} {Reports on Progress in Physics}\ }\textbf
  {\bibinfo {volume} {75}},\ \bibinfo {pages} {076501} (\bibinfo {year}
  {2012})}\BibitemShut {NoStop}%
\bibitem [{\citenamefont {Bergeret}\ \emph {et~al.}(2001)\citenamefont
  {Bergeret}, \citenamefont {Volkov},\ and\ \citenamefont
  {Efetov}}]{bergeret-1}%
  \BibitemOpen
  \bibfield  {author} {\bibinfo {author} {\bibfnamefont {F.~S.}\ \bibnamefont
  {Bergeret}}, \bibinfo {author} {\bibfnamefont {A.~F.}\ \bibnamefont
  {Volkov}}, \ and\ \bibinfo {author} {\bibfnamefont {K.~B.}\ \bibnamefont
  {Efetov}},\ }\href {\doibase 10.1103/PhysRevLett.86.4096} {\bibfield
  {journal} {\bibinfo  {journal} {Phys. Rev. Lett.}\ }\textbf {\bibinfo
  {volume} {86}},\ \bibinfo {pages} {4096} (\bibinfo {year}
  {2001})}\BibitemShut {NoStop}%
\bibitem [{\citenamefont {Ren}\ \emph {et~al.}(2019)\citenamefont {Ren},
  \citenamefont {Pientka}, \citenamefont {Hart}, \citenamefont {Pierce},
  \citenamefont {Kosowsky}, \citenamefont {Lunczer}, \citenamefont {Schlereth},
  \citenamefont {Scharf}, \citenamefont {Hankiewicz}, \citenamefont
  {Molenkamp}, \citenamefont {Halperin},\ and\ \citenamefont
  {Yacoby}}]{Ren2019}%
  \BibitemOpen
  \bibfield  {author} {\bibinfo {author} {\bibfnamefont {H.}~\bibnamefont
  {Ren}}, \bibinfo {author} {\bibfnamefont {F.}~\bibnamefont {Pientka}},
  \bibinfo {author} {\bibfnamefont {S.}~\bibnamefont {Hart}}, \bibinfo {author}
  {\bibfnamefont {A.~T.}\ \bibnamefont {Pierce}}, \bibinfo {author}
  {\bibfnamefont {M.}~\bibnamefont {Kosowsky}}, \bibinfo {author}
  {\bibfnamefont {L.}~\bibnamefont {Lunczer}}, \bibinfo {author} {\bibfnamefont
  {R.}~\bibnamefont {Schlereth}}, \bibinfo {author} {\bibfnamefont
  {B.}~\bibnamefont {Scharf}}, \bibinfo {author} {\bibfnamefont {E.~M.}\
  \bibnamefont {Hankiewicz}}, \bibinfo {author} {\bibfnamefont {L.~W.}\
  \bibnamefont {Molenkamp}}, \bibinfo {author} {\bibfnamefont {B.~I.}\
  \bibnamefont {Halperin}}, \ and\ \bibinfo {author} {\bibfnamefont
  {A.}~\bibnamefont {Yacoby}},\ }\href {\doibase 10.1038/s41586-019-1148-9}
  {\bibfield  {journal} {\bibinfo  {journal} {Nature}\ }\textbf {\bibinfo
  {volume} {569}},\ \bibinfo {pages} {93} (\bibinfo {year} {2019})}\BibitemShut
  {NoStop}%
\bibitem [{\citenamefont {Buzdin}(2005)}]{buzdin-rmp}%
  \BibitemOpen
  \bibfield  {author} {\bibinfo {author} {\bibfnamefont {A.~I.}\ \bibnamefont
  {Buzdin}},\ }\href {\doibase 10.1103/RevModPhys.77.935} {\bibfield  {journal}
  {\bibinfo  {journal} {Rev. Mod. Phys.}\ }\textbf {\bibinfo {volume} {77}},\
  \bibinfo {pages} {935} (\bibinfo {year} {2005})}\BibitemShut {NoStop}%
\bibitem [{\citenamefont {Bergeret}\ \emph {et~al.}(2005)\citenamefont
  {Bergeret}, \citenamefont {Volkov},\ and\ \citenamefont
  {Efetov}}]{odd-triplet-rmp}%
  \BibitemOpen
  \bibfield  {author} {\bibinfo {author} {\bibfnamefont {F.~S.}\ \bibnamefont
  {Bergeret}}, \bibinfo {author} {\bibfnamefont {A.~F.}\ \bibnamefont
  {Volkov}}, \ and\ \bibinfo {author} {\bibfnamefont {K.~B.}\ \bibnamefont
  {Efetov}},\ }\href {\doibase 10.1103/RevModPhys.77.1321} {\bibfield
  {journal} {\bibinfo  {journal} {Rev. Mod. Phys.}\ }\textbf {\bibinfo {volume}
  {77}},\ \bibinfo {pages} {1321} (\bibinfo {year} {2005})}\BibitemShut
  {NoStop}%
\bibitem [{\citenamefont {Eschrig}(2015)}]{Eschrig-review}%
  \BibitemOpen
  \bibfield  {author} {\bibinfo {author} {\bibfnamefont {M.}~\bibnamefont
  {Eschrig}},\ }\href {\doibase 10.1088/0034-4885/78/10/104501} {\bibfield
  {journal} {\bibinfo  {journal} {Reports on Progress in Physics}\ }\textbf
  {\bibinfo {volume} {78}},\ \bibinfo {pages} {104501} (\bibinfo {year}
  {2015})}\BibitemShut {NoStop}%
\bibitem [{\citenamefont {Lemberger}\ \emph {et~al.}(2008)\citenamefont
  {Lemberger}, \citenamefont {Hetel}, \citenamefont {Hauser},\ and\
  \citenamefont {Yang}}]{sf-twocoil}%
  \BibitemOpen
  \bibfield  {author} {\bibinfo {author} {\bibfnamefont {T.~R.}\ \bibnamefont
  {Lemberger}}, \bibinfo {author} {\bibfnamefont {I.}~\bibnamefont {Hetel}},
  \bibinfo {author} {\bibfnamefont {A.~J.}\ \bibnamefont {Hauser}}, \ and\
  \bibinfo {author} {\bibfnamefont {F.~Y.}\ \bibnamefont {Yang}},\ }\href
  {\doibase 10.1063/1.2832318} {\bibfield  {journal} {\bibinfo  {journal}
  {Journal of Applied Physics}\ }\textbf {\bibinfo {volume} {103}} (\bibinfo
  {year} {2008}),\ 10.1063/1.2832318},\ \bibinfo {note} {07C701}\BibitemShut
  {NoStop}%
\bibitem [{\citenamefont {Bae}\ \emph {et~al.}(2019)\citenamefont {Bae},
  \citenamefont {Lee}, \citenamefont {Zhang}, \citenamefont {Takeuchi},\ and\
  \citenamefont {Anlage}}]{anlage-proximity}%
  \BibitemOpen
  \bibfield  {author} {\bibinfo {author} {\bibfnamefont {S.}~\bibnamefont
  {Bae}}, \bibinfo {author} {\bibfnamefont {S.}~\bibnamefont {Lee}}, \bibinfo
  {author} {\bibfnamefont {X.}~\bibnamefont {Zhang}}, \bibinfo {author}
  {\bibfnamefont {I.}~\bibnamefont {Takeuchi}}, \ and\ \bibinfo {author}
  {\bibfnamefont {S.~M.}\ \bibnamefont {Anlage}},\ }\href {\doibase
  10.1103/PhysRevMaterials.3.124803} {\bibfield  {journal} {\bibinfo  {journal}
  {Phys. Rev. Mater.}\ }\textbf {\bibinfo {volume} {3}},\ \bibinfo {pages}
  {124803} (\bibinfo {year} {2019})}\BibitemShut {NoStop}%
\bibitem [{\citenamefont {Di~Bernardo}\ \emph {et~al.}(2015)\citenamefont
  {Di~Bernardo}, \citenamefont {Salman}, \citenamefont {Wang}, \citenamefont
  {Amado}, \citenamefont {Egilmez}, \citenamefont {Flokstra}, \citenamefont
  {Suter}, \citenamefont {Lee}, \citenamefont {Zhao}, \citenamefont {Prokscha},
  \citenamefont {Morenzoni}, \citenamefont {Blamire}, \citenamefont {Linder},\
  and\ \citenamefont {Robinson}}]{para-meissner-musr}%
  \BibitemOpen
  \bibfield  {author} {\bibinfo {author} {\bibfnamefont {A.}~\bibnamefont
  {Di~Bernardo}}, \bibinfo {author} {\bibfnamefont {Z.}~\bibnamefont {Salman}},
  \bibinfo {author} {\bibfnamefont {X.~L.}\ \bibnamefont {Wang}}, \bibinfo
  {author} {\bibfnamefont {M.}~\bibnamefont {Amado}}, \bibinfo {author}
  {\bibfnamefont {M.}~\bibnamefont {Egilmez}}, \bibinfo {author} {\bibfnamefont
  {M.~G.}\ \bibnamefont {Flokstra}}, \bibinfo {author} {\bibfnamefont
  {A.}~\bibnamefont {Suter}}, \bibinfo {author} {\bibfnamefont {S.~L.}\
  \bibnamefont {Lee}}, \bibinfo {author} {\bibfnamefont {J.~H.}\ \bibnamefont
  {Zhao}}, \bibinfo {author} {\bibfnamefont {T.}~\bibnamefont {Prokscha}},
  \bibinfo {author} {\bibfnamefont {E.}~\bibnamefont {Morenzoni}}, \bibinfo
  {author} {\bibfnamefont {M.~G.}\ \bibnamefont {Blamire}}, \bibinfo {author}
  {\bibfnamefont {J.}~\bibnamefont {Linder}}, \ and\ \bibinfo {author}
  {\bibfnamefont {J.~W.~A.}\ \bibnamefont {Robinson}},\ }\href {\doibase
  10.1103/PhysRevX.5.041021} {\bibfield  {journal} {\bibinfo  {journal} {Phys.
  Rev. X}\ }\textbf {\bibinfo {volume} {5}},\ \bibinfo {pages} {041021}
  (\bibinfo {year} {2015})}\BibitemShut {NoStop}%
\bibitem [{\citenamefont {Phan}\ \emph {et~al.}(2022)\citenamefont {Phan},
  \citenamefont {Senior}, \citenamefont {Ghazaryan}, \citenamefont
  {Hatefipour}, \citenamefont {Strickland}, \citenamefont {Shabani},
  \citenamefont {Serbyn},\ and\ \citenamefont
  {Higginbotham}}]{higginbotham-prl}%
  \BibitemOpen
  \bibfield  {author} {\bibinfo {author} {\bibfnamefont {D.}~\bibnamefont
  {Phan}}, \bibinfo {author} {\bibfnamefont {J.}~\bibnamefont {Senior}},
  \bibinfo {author} {\bibfnamefont {A.}~\bibnamefont {Ghazaryan}}, \bibinfo
  {author} {\bibfnamefont {M.}~\bibnamefont {Hatefipour}}, \bibinfo {author}
  {\bibfnamefont {W.~M.}\ \bibnamefont {Strickland}}, \bibinfo {author}
  {\bibfnamefont {J.}~\bibnamefont {Shabani}}, \bibinfo {author} {\bibfnamefont
  {M.}~\bibnamefont {Serbyn}}, \ and\ \bibinfo {author} {\bibfnamefont {A.~P.}\
  \bibnamefont {Higginbotham}},\ }\href {\doibase
  10.1103/PhysRevLett.128.107701} {\bibfield  {journal} {\bibinfo  {journal}
  {Phys. Rev. Lett.}\ }\textbf {\bibinfo {volume} {128}},\ \bibinfo {pages}
  {107701} (\bibinfo {year} {2022})}\BibitemShut {NoStop}%
\bibitem [{\citenamefont {Han}\ \emph {et~al.}(2020)\citenamefont {Han},
  \citenamefont {Maekawa},\ and\ \citenamefont {Xie}}]{spincurrent}%
  \BibitemOpen
  \bibfield  {author} {\bibinfo {author} {\bibfnamefont {W.}~\bibnamefont
  {Han}}, \bibinfo {author} {\bibfnamefont {S.}~\bibnamefont {Maekawa}}, \ and\
  \bibinfo {author} {\bibfnamefont {X.-C.}\ \bibnamefont {Xie}},\ }\href
  {\doibase 10.1038/s41563-019-0456-7} {\bibfield  {journal} {\bibinfo
  {journal} {Nature Materials}\ }\textbf {\bibinfo {volume} {19}},\ \bibinfo
  {pages} {139} (\bibinfo {year} {2020})}\BibitemShut {NoStop}%
\bibitem [{\citenamefont {Tabuchi}\ \emph {et~al.}(2014)\citenamefont
  {Tabuchi}, \citenamefont {Ishino}, \citenamefont {Ishikawa}, \citenamefont
  {Yamazaki}, \citenamefont {Usami},\ and\ \citenamefont {Nakamura}}]{yig-3d}%
  \BibitemOpen
  \bibfield  {author} {\bibinfo {author} {\bibfnamefont {Y.}~\bibnamefont
  {Tabuchi}}, \bibinfo {author} {\bibfnamefont {S.}~\bibnamefont {Ishino}},
  \bibinfo {author} {\bibfnamefont {T.}~\bibnamefont {Ishikawa}}, \bibinfo
  {author} {\bibfnamefont {R.}~\bibnamefont {Yamazaki}}, \bibinfo {author}
  {\bibfnamefont {K.}~\bibnamefont {Usami}}, \ and\ \bibinfo {author}
  {\bibfnamefont {Y.}~\bibnamefont {Nakamura}},\ }\href {\doibase
  10.1103/PhysRevLett.113.083603} {\bibfield  {journal} {\bibinfo  {journal}
  {Phys. Rev. Lett.}\ }\textbf {\bibinfo {volume} {113}},\ \bibinfo {pages}
  {083603} (\bibinfo {year} {2014})}\BibitemShut {NoStop}%
\bibitem [{\citenamefont {Zhang}\ \emph {et~al.}(2014)\citenamefont {Zhang},
  \citenamefont {Zou}, \citenamefont {Jiang},\ and\ \citenamefont
  {Tang}}]{yig-3d-2}%
  \BibitemOpen
  \bibfield  {author} {\bibinfo {author} {\bibfnamefont {X.}~\bibnamefont
  {Zhang}}, \bibinfo {author} {\bibfnamefont {C.-L.}\ \bibnamefont {Zou}},
  \bibinfo {author} {\bibfnamefont {L.}~\bibnamefont {Jiang}}, \ and\ \bibinfo
  {author} {\bibfnamefont {H.~X.}\ \bibnamefont {Tang}},\ }\href {\doibase
  10.1103/PhysRevLett.113.156401} {\bibfield  {journal} {\bibinfo  {journal}
  {Phys. Rev. Lett.}\ }\textbf {\bibinfo {volume} {113}},\ \bibinfo {pages}
  {156401} (\bibinfo {year} {2014})}\BibitemShut {NoStop}%
\bibitem [{\citenamefont {Huebl}\ \emph {et~al.}(2013)\citenamefont {Huebl},
  \citenamefont {Zollitsch}, \citenamefont {Lotze}, \citenamefont {Hocke},
  \citenamefont {Greifenstein}, \citenamefont {Marx}, \citenamefont {Gross},\
  and\ \citenamefont {Goennenwein}}]{yig-2d}%
  \BibitemOpen
  \bibfield  {author} {\bibinfo {author} {\bibfnamefont {H.}~\bibnamefont
  {Huebl}}, \bibinfo {author} {\bibfnamefont {C.~W.}\ \bibnamefont
  {Zollitsch}}, \bibinfo {author} {\bibfnamefont {J.}~\bibnamefont {Lotze}},
  \bibinfo {author} {\bibfnamefont {F.}~\bibnamefont {Hocke}}, \bibinfo
  {author} {\bibfnamefont {M.}~\bibnamefont {Greifenstein}}, \bibinfo {author}
  {\bibfnamefont {A.}~\bibnamefont {Marx}}, \bibinfo {author} {\bibfnamefont
  {R.}~\bibnamefont {Gross}}, \ and\ \bibinfo {author} {\bibfnamefont
  {S.~T.~B.}\ \bibnamefont {Goennenwein}},\ }\href {\doibase
  10.1103/PhysRevLett.111.127003} {\bibfield  {journal} {\bibinfo  {journal}
  {Phys. Rev. Lett.}\ }\textbf {\bibinfo {volume} {111}},\ \bibinfo {pages}
  {127003} (\bibinfo {year} {2013})}\BibitemShut {NoStop}%
\bibitem [{\citenamefont {Hou}\ and\ \citenamefont {Liu}(2019)}]{nb-py-prl1}%
  \BibitemOpen
  \bibfield  {author} {\bibinfo {author} {\bibfnamefont {J.~T.}\ \bibnamefont
  {Hou}}\ and\ \bibinfo {author} {\bibfnamefont {L.}~\bibnamefont {Liu}},\
  }\href {\doibase 10.1103/PhysRevLett.123.107702} {\bibfield  {journal}
  {\bibinfo  {journal} {Phys. Rev. Lett.}\ }\textbf {\bibinfo {volume} {123}},\
  \bibinfo {pages} {107702} (\bibinfo {year} {2019})}\BibitemShut {NoStop}%
\bibitem [{\citenamefont {Li}\ \emph {et~al.}(2019)\citenamefont {Li},
  \citenamefont {Polakovic}, \citenamefont {Wang}, \citenamefont {Xu},
  \citenamefont {Lendinez}, \citenamefont {Zhang}, \citenamefont {Ding},
  \citenamefont {Khaire}, \citenamefont {Saglam}, \citenamefont {Divan},
  \citenamefont {Pearson}, \citenamefont {Kwok}, \citenamefont {Xiao},
  \citenamefont {Novosad}, \citenamefont {Hoffmann},\ and\ \citenamefont
  {Zhang}}]{nb-py-prl2}%
  \BibitemOpen
  \bibfield  {author} {\bibinfo {author} {\bibfnamefont {Y.}~\bibnamefont
  {Li}}, \bibinfo {author} {\bibfnamefont {T.}~\bibnamefont {Polakovic}},
  \bibinfo {author} {\bibfnamefont {Y.-L.}\ \bibnamefont {Wang}}, \bibinfo
  {author} {\bibfnamefont {J.}~\bibnamefont {Xu}}, \bibinfo {author}
  {\bibfnamefont {S.}~\bibnamefont {Lendinez}}, \bibinfo {author}
  {\bibfnamefont {Z.}~\bibnamefont {Zhang}}, \bibinfo {author} {\bibfnamefont
  {J.}~\bibnamefont {Ding}}, \bibinfo {author} {\bibfnamefont {T.}~\bibnamefont
  {Khaire}}, \bibinfo {author} {\bibfnamefont {H.}~\bibnamefont {Saglam}},
  \bibinfo {author} {\bibfnamefont {R.}~\bibnamefont {Divan}}, \bibinfo
  {author} {\bibfnamefont {J.}~\bibnamefont {Pearson}}, \bibinfo {author}
  {\bibfnamefont {W.-K.}\ \bibnamefont {Kwok}}, \bibinfo {author}
  {\bibfnamefont {Z.}~\bibnamefont {Xiao}}, \bibinfo {author} {\bibfnamefont
  {V.}~\bibnamefont {Novosad}}, \bibinfo {author} {\bibfnamefont
  {A.}~\bibnamefont {Hoffmann}}, \ and\ \bibinfo {author} {\bibfnamefont
  {W.}~\bibnamefont {Zhang}},\ }\href {\doibase 10.1103/PhysRevLett.123.107701}
  {\bibfield  {journal} {\bibinfo  {journal} {Phys. Rev. Lett.}\ }\textbf
  {\bibinfo {volume} {123}},\ \bibinfo {pages} {107701} (\bibinfo {year}
  {2019})}\BibitemShut {NoStop}%
\bibitem [{\citenamefont {Prozorov}\ and\ \citenamefont
  {Giannetta}(2006)}]{Prozorov}%
  \BibitemOpen
  \bibfield  {author} {\bibinfo {author} {\bibfnamefont {R.}~\bibnamefont
  {Prozorov}}\ and\ \bibinfo {author} {\bibfnamefont {R.~W.}\ \bibnamefont
  {Giannetta}},\ }\href {\doibase 10.1088/0953-2048/19/8/r01} {\bibfield
  {journal} {\bibinfo  {journal} {Superconductor Science and Technology}\
  }\textbf {\bibinfo {volume} {19}},\ \bibinfo {pages} {R41} (\bibinfo {year}
  {2006})}\BibitemShut {NoStop}%
\bibitem [{\citenamefont {Demler}\ \emph {et~al.}(1997)\citenamefont {Demler},
  \citenamefont {Arnold},\ and\ \citenamefont {Beasley}}]{eugene-sf}%
  \BibitemOpen
  \bibfield  {author} {\bibinfo {author} {\bibfnamefont {E.~A.}\ \bibnamefont
  {Demler}}, \bibinfo {author} {\bibfnamefont {G.~B.}\ \bibnamefont {Arnold}},
  \ and\ \bibinfo {author} {\bibfnamefont {M.~R.}\ \bibnamefont {Beasley}},\
  }\href {\doibase 10.1103/PhysRevB.55.15174} {\bibfield  {journal} {\bibinfo
  {journal} {Phys. Rev. B}\ }\textbf {\bibinfo {volume} {55}},\ \bibinfo
  {pages} {15174} (\bibinfo {year} {1997})}\BibitemShut {NoStop}%
\bibitem [{\citenamefont {Bergeret}\ and\ \citenamefont
  {Tokatly}(2013)}]{sf-soc}%
  \BibitemOpen
  \bibfield  {author} {\bibinfo {author} {\bibfnamefont {F.~S.}\ \bibnamefont
  {Bergeret}}\ and\ \bibinfo {author} {\bibfnamefont {I.~V.}\ \bibnamefont
  {Tokatly}},\ }\href {\doibase 10.1103/PhysRevLett.110.117003} {\bibfield
  {journal} {\bibinfo  {journal} {Phys. Rev. Lett.}\ }\textbf {\bibinfo
  {volume} {110}},\ \bibinfo {pages} {117003} (\bibinfo {year}
  {2013})}\BibitemShut {NoStop}%
\bibitem [{\citenamefont {Takei}\ and\ \citenamefont
  {Galitski}(2012)}]{takei_SF}%
  \BibitemOpen
  \bibfield  {author} {\bibinfo {author} {\bibfnamefont {S.}~\bibnamefont
  {Takei}}\ and\ \bibinfo {author} {\bibfnamefont {V.}~\bibnamefont
  {Galitski}},\ }\href@noop {} {\bibfield  {journal} {\bibinfo  {journal}
  {Physical Review B}\ }\textbf {\bibinfo {volume} {86}},\ \bibinfo {pages}
  {054521} (\bibinfo {year} {2012})}\BibitemShut {NoStop}%
\bibitem [{\citenamefont {Linder}\ and\ \citenamefont
  {Balatsky}(2019)}]{odd-w-rmp}%
  \BibitemOpen
  \bibfield  {author} {\bibinfo {author} {\bibfnamefont {J.}~\bibnamefont
  {Linder}}\ and\ \bibinfo {author} {\bibfnamefont {A.~V.}\ \bibnamefont
  {Balatsky}},\ }\href {\doibase 10.1103/RevModPhys.91.045005} {\bibfield
  {journal} {\bibinfo  {journal} {Rev. Mod. Phys.}\ }\textbf {\bibinfo {volume}
  {91}},\ \bibinfo {pages} {045005} (\bibinfo {year} {2019})}\BibitemShut
  {NoStop}%
\bibitem [{\citenamefont {Robinson}\ \emph {et~al.}(2010)\citenamefont
  {Robinson}, \citenamefont {Witt},\ and\ \citenamefont {Blamire}}]{sfs-jj-1}%
  \BibitemOpen
  \bibfield  {author} {\bibinfo {author} {\bibfnamefont {J.~W.~A.}\
  \bibnamefont {Robinson}}, \bibinfo {author} {\bibfnamefont {J.~D.~S.}\
  \bibnamefont {Witt}}, \ and\ \bibinfo {author} {\bibfnamefont {M.~G.}\
  \bibnamefont {Blamire}},\ }\href {\doibase 10.1126/science.1189246}
  {\bibfield  {journal} {\bibinfo  {journal} {Science}\ }\textbf {\bibinfo
  {volume} {329}},\ \bibinfo {pages} {59} (\bibinfo {year} {2010})}\BibitemShut
  {NoStop}%
\bibitem [{\citenamefont {Khaire}\ \emph {et~al.}(2010)\citenamefont {Khaire},
  \citenamefont {Khasawneh}, \citenamefont {Pratt},\ and\ \citenamefont
  {Birge}}]{sfs-jj-2}%
  \BibitemOpen
  \bibfield  {author} {\bibinfo {author} {\bibfnamefont {T.~S.}\ \bibnamefont
  {Khaire}}, \bibinfo {author} {\bibfnamefont {M.~A.}\ \bibnamefont
  {Khasawneh}}, \bibinfo {author} {\bibfnamefont {W.~P.}\ \bibnamefont
  {Pratt}}, \ and\ \bibinfo {author} {\bibfnamefont {N.~O.}\ \bibnamefont
  {Birge}},\ }\href {\doibase 10.1103/PhysRevLett.104.137002} {\bibfield
  {journal} {\bibinfo  {journal} {Phys. Rev. Lett.}\ }\textbf {\bibinfo
  {volume} {104}},\ \bibinfo {pages} {137002} (\bibinfo {year}
  {2010})}\BibitemShut {NoStop}%
\bibitem [{\citenamefont {Keizer}\ \emph {et~al.}(2006)\citenamefont {Keizer},
  \citenamefont {Goennenwein}, \citenamefont {Klapwijk}, \citenamefont {Miao},
  \citenamefont {Xiao},\ and\ \citenamefont {Gupta}}]{sfs-jj-3}%
  \BibitemOpen
  \bibfield  {author} {\bibinfo {author} {\bibfnamefont {R.~S.}\ \bibnamefont
  {Keizer}}, \bibinfo {author} {\bibfnamefont {S.~T.~B.}\ \bibnamefont
  {Goennenwein}}, \bibinfo {author} {\bibfnamefont {T.~M.}\ \bibnamefont
  {Klapwijk}}, \bibinfo {author} {\bibfnamefont {G.}~\bibnamefont {Miao}},
  \bibinfo {author} {\bibfnamefont {G.}~\bibnamefont {Xiao}}, \ and\ \bibinfo
  {author} {\bibfnamefont {A.}~\bibnamefont {Gupta}},\ }\href {\doibase
  10.1038/nature04499} {\bibfield  {journal} {\bibinfo  {journal} {Nature}\
  }\textbf {\bibinfo {volume} {439}},\ \bibinfo {pages} {825} (\bibinfo {year}
  {2006})}\BibitemShut {NoStop}%
\bibitem [{\citenamefont {Blais}\ \emph {et~al.}(2021)\citenamefont {Blais},
  \citenamefont {Grimsmo}, \citenamefont {Girvin},\ and\ \citenamefont
  {Wallraff}}]{blais-rmp}%
  \BibitemOpen
  \bibfield  {author} {\bibinfo {author} {\bibfnamefont {A.}~\bibnamefont
  {Blais}}, \bibinfo {author} {\bibfnamefont {A.~L.}\ \bibnamefont {Grimsmo}},
  \bibinfo {author} {\bibfnamefont {S.~M.}\ \bibnamefont {Girvin}}, \ and\
  \bibinfo {author} {\bibfnamefont {A.}~\bibnamefont {Wallraff}},\ }\href
  {\doibase 10.1103/RevModPhys.93.025005} {\bibfield  {journal} {\bibinfo
  {journal} {Rev. Mod. Phys.}\ }\textbf {\bibinfo {volume} {93}},\ \bibinfo
  {pages} {025005} (\bibinfo {year} {2021})}\BibitemShut {NoStop}%
\bibitem [{\citenamefont {Kroll}\ \emph {et~al.}(2019)\citenamefont {Kroll},
  \citenamefont {Borsoi}, \citenamefont {van~der Enden}, \citenamefont
  {Uilhoorn}, \citenamefont {de~Jong}, \citenamefont {Quintero-P\'erez},
  \citenamefont {van Woerkom}, \citenamefont {Bruno}, \citenamefont {Plissard},
  \citenamefont {Car}, \citenamefont {Bakkers}, \citenamefont {Cassidy},\ and\
  \citenamefont {Kouwenhoven}}]{leo-fluxholes}%
  \BibitemOpen
  \bibfield  {author} {\bibinfo {author} {\bibfnamefont {J.}~\bibnamefont
  {Kroll}}, \bibinfo {author} {\bibfnamefont {F.}~\bibnamefont {Borsoi}},
  \bibinfo {author} {\bibfnamefont {K.}~\bibnamefont {van~der Enden}}, \bibinfo
  {author} {\bibfnamefont {W.}~\bibnamefont {Uilhoorn}}, \bibinfo {author}
  {\bibfnamefont {D.}~\bibnamefont {de~Jong}}, \bibinfo {author} {\bibfnamefont
  {M.}~\bibnamefont {Quintero-P\'erez}}, \bibinfo {author} {\bibfnamefont
  {D.}~\bibnamefont {van Woerkom}}, \bibinfo {author} {\bibfnamefont
  {A.}~\bibnamefont {Bruno}}, \bibinfo {author} {\bibfnamefont
  {S.}~\bibnamefont {Plissard}}, \bibinfo {author} {\bibfnamefont
  {D.}~\bibnamefont {Car}}, \bibinfo {author} {\bibfnamefont {E.}~\bibnamefont
  {Bakkers}}, \bibinfo {author} {\bibfnamefont {M.}~\bibnamefont {Cassidy}}, \
  and\ \bibinfo {author} {\bibfnamefont {L.}~\bibnamefont {Kouwenhoven}},\
  }\href {\doibase 10.1103/PhysRevApplied.11.064053} {\bibfield  {journal}
  {\bibinfo  {journal} {Phys. Rev. Applied}\ }\textbf {\bibinfo {volume}
  {11}},\ \bibinfo {pages} {064053} (\bibinfo {year} {2019})}\BibitemShut
  {NoStop}%
\bibitem [{\citenamefont {Kittel}(1948)}]{kittel}%
  \BibitemOpen
  \bibfield  {author} {\bibinfo {author} {\bibfnamefont {C.}~\bibnamefont
  {Kittel}},\ }\href {\doibase 10.1103/PhysRev.73.155} {\bibfield  {journal}
  {\bibinfo  {journal} {Phys. Rev.}\ }\textbf {\bibinfo {volume} {73}},\
  \bibinfo {pages} {155} (\bibinfo {year} {1948})}\BibitemShut {NoStop}%
\bibitem [{\citenamefont {Prabhakar}\ and\ \citenamefont
  {Stancil}(2009)}]{sw-book}%
  \BibitemOpen
  \bibfield  {author} {\bibinfo {author} {\bibfnamefont {A.}~\bibnamefont
  {Prabhakar}}\ and\ \bibinfo {author} {\bibfnamefont {D.}~\bibnamefont
  {Stancil}},\ }\href@noop {} {\emph {\bibinfo {title} {Spin Waves: Theory and
  Applications}}}\ (\bibinfo  {publisher} {Springer},\ \bibinfo {address}
  {Boston},\ \bibinfo {year} {2009})\BibitemShut {NoStop}%
\bibitem [{\citenamefont {de~Gennes}(1966)}]{degennes-book}%
  \BibitemOpen
  \bibfield  {author} {\bibinfo {author} {\bibfnamefont {P.~G.}\ \bibnamefont
  {de~Gennes}},\ }\href@noop {} {\emph {\bibinfo {title} {Superconductivity of
  Metals and Alloys}}}\ (\bibinfo  {publisher} {Benjamin},\ \bibinfo {address}
  {New York},\ \bibinfo {year} {1966})\BibitemShut {NoStop}%
\bibitem [{\citenamefont {Sigrist}\ and\ \citenamefont
  {Ueda}(1991)}]{sigrist-rmp}%
  \BibitemOpen
  \bibfield  {author} {\bibinfo {author} {\bibfnamefont {M.}~\bibnamefont
  {Sigrist}}\ and\ \bibinfo {author} {\bibfnamefont {K.}~\bibnamefont {Ueda}},\
  }\href {\doibase 10.1103/RevModPhys.63.239} {\bibfield  {journal} {\bibinfo
  {journal} {Rev. Mod. Phys.}\ }\textbf {\bibinfo {volume} {63}},\ \bibinfo
  {pages} {239} (\bibinfo {year} {1991})}\BibitemShut {NoStop}%
\bibitem [{\citenamefont {Leggett}(1975)}]{leggett-rmp}%
  \BibitemOpen
  \bibfield  {author} {\bibinfo {author} {\bibfnamefont {A.~J.}\ \bibnamefont
  {Leggett}},\ }\href {\doibase 10.1103/RevModPhys.47.331} {\bibfield
  {journal} {\bibinfo  {journal} {Rev. Mod. Phys.}\ }\textbf {\bibinfo {volume}
  {47}},\ \bibinfo {pages} {331} (\bibinfo {year} {1975})}\BibitemShut
  {NoStop}%
\bibitem [{\citenamefont {Vollhardt}\ and\ \citenamefont
  {Wolfle}(1990)}]{3he-book}%
  \BibitemOpen
  \bibfield  {author} {\bibinfo {author} {\bibfnamefont {D.}~\bibnamefont
  {Vollhardt}}\ and\ \bibinfo {author} {\bibfnamefont {P.}~\bibnamefont
  {Wolfle}},\ }\href@noop {} {\emph {\bibinfo {title} {The superfluid phases of
  {h}elium 3}}}\ (\bibinfo  {publisher} {Taylor and Francis},\ \bibinfo
  {address} {London},\ \bibinfo {year} {1990})\BibitemShut {NoStop}%
\bibitem [{\citenamefont {Hirschfeld}\ and\ \citenamefont
  {Goldenfeld}(1993)}]{hirschfeld-disorder}%
  \BibitemOpen
  \bibfield  {author} {\bibinfo {author} {\bibfnamefont {P.~J.}\ \bibnamefont
  {Hirschfeld}}\ and\ \bibinfo {author} {\bibfnamefont {N.}~\bibnamefont
  {Goldenfeld}},\ }\href {\doibase 10.1103/PhysRevB.48.4219} {\bibfield
  {journal} {\bibinfo  {journal} {Phys. Rev. B}\ }\textbf {\bibinfo {volume}
  {48}},\ \bibinfo {pages} {4219} (\bibinfo {year} {1993})}\BibitemShut
  {NoStop}%
\bibitem [{\citenamefont {B{\o}ttcher}(2022)}]{cboettcher2022}%
  \BibitemOpen
  \bibfield  {author} {\bibinfo {author} {\bibfnamefont {C.~G.~L.}\
  \bibnamefont {B{\o}ttcher}},\ }\emph {\bibinfo {title} {New avenues in
  circuit QED: from quantum information to quantum sensing}},\ \href@noop {}
  {Ph.D. thesis},\ \bibinfo  {school} {Harvard University Graduate School of
  Arts and Sciences} (\bibinfo {year} {2022})\BibitemShut {NoStop}%
\bibitem [{\citenamefont {Mahan}(2000)}]{mahan_txtbk}%
  \BibitemOpen
  \bibfield  {author} {\bibinfo {author} {\bibfnamefont {G.~D.}\ \bibnamefont
  {Mahan}},\ }\href@noop {} {\emph {\bibinfo {title} {Many-particle physics}}}\
  (\bibinfo  {publisher} {Springer Science \& Business Media},\ \bibinfo {year}
  {2000})\BibitemShut {NoStop}%
\bibitem [{\citenamefont {Chen}\ \emph {et~al.}(2022)\citenamefont {Chen},
  \citenamefont {Pfeiffer}, \citenamefont {Partanen}, \citenamefont {Fesquet},
  \citenamefont {Honasoge}, \citenamefont {Kronowetter}, \citenamefont
  {Nojiri}, \citenamefont {Renger}, \citenamefont {Fedorov}, \citenamefont
  {Marx}, \citenamefont {Deppe},\ and\ \citenamefont {Gross}}]{hanging-res}%
  \BibitemOpen
  \bibfield  {author} {\bibinfo {author} {\bibfnamefont {Q.-M.}\ \bibnamefont
  {Chen}}, \bibinfo {author} {\bibfnamefont {M.}~\bibnamefont {Pfeiffer}},
  \bibinfo {author} {\bibfnamefont {M.}~\bibnamefont {Partanen}}, \bibinfo
  {author} {\bibfnamefont {F.}~\bibnamefont {Fesquet}}, \bibinfo {author}
  {\bibfnamefont {K.~E.}\ \bibnamefont {Honasoge}}, \bibinfo {author}
  {\bibfnamefont {F.}~\bibnamefont {Kronowetter}}, \bibinfo {author}
  {\bibfnamefont {Y.}~\bibnamefont {Nojiri}}, \bibinfo {author} {\bibfnamefont
  {M.}~\bibnamefont {Renger}}, \bibinfo {author} {\bibfnamefont {K.~G.}\
  \bibnamefont {Fedorov}}, \bibinfo {author} {\bibfnamefont {A.}~\bibnamefont
  {Marx}}, \bibinfo {author} {\bibfnamefont {F.}~\bibnamefont {Deppe}}, \ and\
  \bibinfo {author} {\bibfnamefont {R.}~\bibnamefont {Gross}},\ }\href
  {\doibase 10.1103/PhysRevB.106.214505} {\bibfield  {journal} {\bibinfo
  {journal} {Phys. Rev. B}\ }\textbf {\bibinfo {volume} {106}},\ \bibinfo
  {pages} {214505} (\bibinfo {year} {2022})}\BibitemShut {NoStop}%
\bibitem [{\citenamefont {Makita}\ \emph {et~al.}(2022)\citenamefont {Makita},
  \citenamefont {Sundahl}, \citenamefont {Ciovati}, \citenamefont {Eom},\ and\
  \citenamefont {Gurevich}}]{nlme-cpw}%
  \BibitemOpen
  \bibfield  {author} {\bibinfo {author} {\bibfnamefont {J.}~\bibnamefont
  {Makita}}, \bibinfo {author} {\bibfnamefont {C.}~\bibnamefont {Sundahl}},
  \bibinfo {author} {\bibfnamefont {G.}~\bibnamefont {Ciovati}}, \bibinfo
  {author} {\bibfnamefont {C.~B.}\ \bibnamefont {Eom}}, \ and\ \bibinfo
  {author} {\bibfnamefont {A.}~\bibnamefont {Gurevich}},\ }\href {\doibase
  10.1103/PhysRevResearch.4.013156} {\bibfield  {journal} {\bibinfo  {journal}
  {Phys. Rev. Res.}\ }\textbf {\bibinfo {volume} {4}},\ \bibinfo {pages}
  {013156} (\bibinfo {year} {2022})}\BibitemShut {NoStop}%
\end{thebibliography}%

\end{document}